%% file: IMWUT_25/main.tex
\documentclass[acmlarge]{acmart}

\acmJournal{IMWUT}
\acmYear{2025} 
\acmVolume{9} 
\acmNumber{3} 
\acmArticle{95} 
\acmMonth{9} 
\acmPrice{}\acmDOI{10.1145/3749498}

\usepackage{hyperref}
\usepackage{url}
\usepackage{derivative}
\usepackage{soul}
\usepackage{graphicx}
\usepackage{amsmath}
\usepackage{mathtools}
\usepackage{xspace}
\usepackage{booktabs}
\usepackage{xcolor}
\usepackage{color, colortbl}
\usepackage{diagbox}
\usepackage{multirow}
\usepackage{multicol}
\usepackage{lipsum}
\usepackage{tabularray}
\usepackage{colortbl}
\usepackage{enumitem}

\usepackage{subcaption}
\usepackage{parcolumns}
\usepackage{listings}
\usepackage{array}
\usepackage{tabularx,ragged2e}

\definecolor{armygreen}{rgb}{0.29, 0.33, 0.13}
\definecolor{darkgreen}{rgb}{0.0, 0.5, 0.0}

\newcommand{\parlabel}[1]{\vspace{0.5em}\noindent\textbf{#1}.}

\begin{document}

\title{Mindfulness Meditation and Respiration: Accelerometer-Based Respiration Rate and Mindfulness Progress Estimation to Enhance App Engagement and Mindfulness Skills}

\input{IMWUT_25/tex/00_abstract}



\received{1 February 2025}
\received[revised]{1 May 2025}
\received[accepted]{5 July 2025}



\author{Mohammad Nur Hossain Khan}
\email{mkhan@wpi.edu}
\affiliation{%
\institution{Worcester Polytechnic Institute}
\city{Worcester}
  \state{Massachusetts}
  \country{USA}
}

\author{David creswell}
\email{creswell@andrew.cmu.edu}
\affiliation{%
\institution{Carnegie Mellon University}
\city{Pittsburgh}
  \state{Pennsylvania}
  \country{USA}
}
\affiliation{%
\institution{Equa Health Inc.}
\city{Pittsburgh}
  \state{Pennsylvania}
  \country{USA}
}

\author{Jordan Albert}
\email{jordan@equahealth.io}
\affiliation{%
\institution{Carnegie Mellon University}
\city{Pittsburgh}
  \state{Pennsylvania}
  \country{USA}
}
\affiliation{%
\institution{Equa Health Inc.}
\city{Pittsburgh}
  \state{Pennsylvania}
  \country{USA}
}

\author{Patrick O'Connell}
\email{poconnel@andrew.cmu.edu}
\affiliation{%
\institution{Carnegie Mellon University}
\city{Pittsburgh}
  \state{Pennsylvania}
  \country{USA}
}
\affiliation{%
\institution{Equa Health Inc.}
\city{Pittsburgh}
  \state{Pennsylvania}
  \country{USA}
}

\author{Shawn Fallon}
\email{shawn@equahealth.io}
\affiliation{%
\institution{Carnegie Mellon University}
\city{Pittsburgh}
  \state{Pennsylvania}
  \country{USA}
}
\affiliation{%
\institution{Equa Health Inc.}
\city{Pittsburgh}
  \state{Pennsylvania}
  \country{USA}
}

\author{Mathew Polowitz}
\email{mathew@equahealth.io}
\affiliation{%
\institution{Carnegie Mellon University}
\city{Pittsburgh}
  \state{Pennsylvania}
  \country{USA}
}
\affiliation{%
\institution{Equa Health Inc.}
\city{Pittsburgh}
  \state{Pennsylvania}
  \country{USA}
}

\author{Xuhai "orson" Xu}
\email{xx2489@cumc.columbia.edu}
\affiliation{%
\institution{Columbia University}
\city{New York}
  \state{new York}
  \country{USA}
}

\author{Bashima islam}
\email{bislam@wpi.edu}
\affiliation{%
\institution{Worcester Polytechnic Institute}
\city{Worcester}
  \state{Massachusetts}
  \country{USA}
}


\maketitle
\input{IMWUT_25/tex/01_introduction}

\input{IMWUT_25/tex/background}

\input{IMWUT_25/tex/03_method}
\input{IMWUT_25/tex/05_result}
\input{IMWUT_25/tex/06_discussion}

\input{IMWUT_25/tex/07_conclusion}

\bibliographystyle{ACM-Reference-Format}
\bibliography{IMWUT_25/reference}

\end{document}

%% file: IMWUT_25/tex/00_abstract.tex
\begin{abstract}
Mindfulness training is widely recognized for its benefits in reducing depression, anxiety, and loneliness. With the rise of smartphone-based mindfulness apps, digital meditation has become more accessible, but sustaining long-term user engagement remains a challenge. This paper explores whether respiration biosignal feedback and mindfulness skill estimation enhance system usability and skill development. We develop a smartphone’s accelerometer-based respiration tracking algorithm, eliminating the need for additional wearables. Unlike existing methods, our approach accurately captures slow breathing patterns typical of mindfulness meditation. Additionally, we introduce the first quantitative framework to estimate mindfulness skills—concentration, sensory clarity, and equanimity—based on accelerometer-derived respiration data. We develop and test our algorithms on 261 mindfulness sessions in both controlled and real-world settings. A user study comparing an experimental group receiving biosignal feedback with a control group using a standard app shows that respiration feedback enhances system usability. Our respiration tracking model achieves a mean absolute error (MAE) of 1.6 breaths per minute, closely aligning with ground truth data, while our mindfulness skill estimation attains F1 scores of 80–84\% in tracking skill progression. By integrating respiration tracking and mindfulness estimation into a commercial app, we demonstrate the potential of smartphone sensors to enhance digital mindfulness training.

\end{abstract}

%% file: IMWUT_25/tex/01_introduction.tex
\begin{figure}
    \centering
    \includegraphics[width=\linewidth]{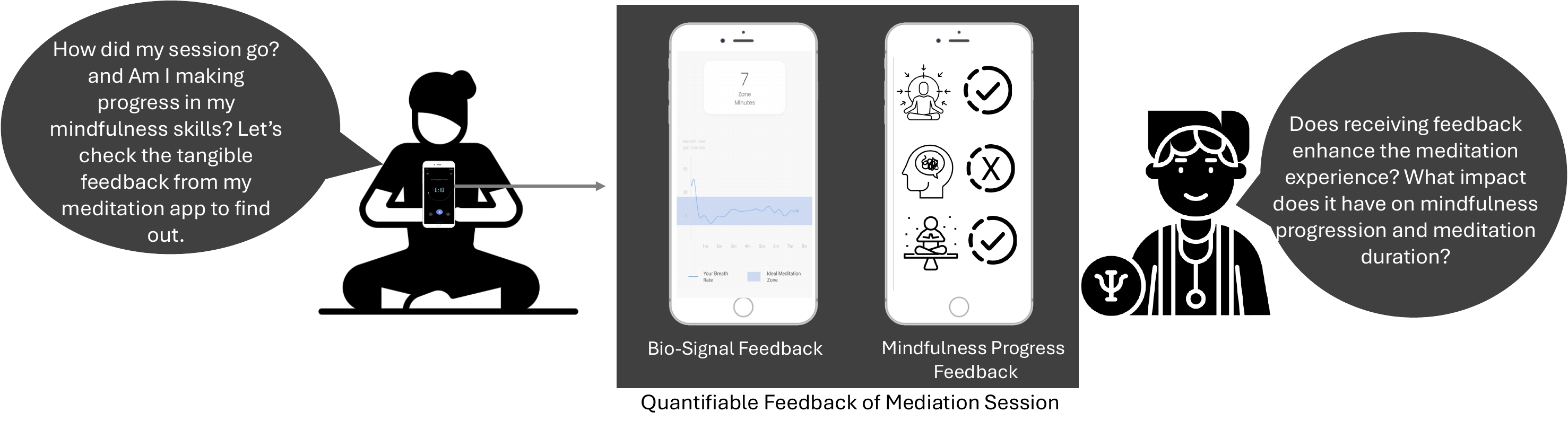}
    \caption{We integrate a feedback module into a commercially available mindfulness meditation app, Equa, featuring a real-time respiration rate chart, respiration statistics, and mindfulness progress feedback. This is achieved by developing a respiration rate tracking algorithm and a mindfulness skill change estimation algorithm. To assess the impact of this feedback, we conducted a user study evaluating user satisfaction, mindfulness progression, and engagement.}
    \label{fig:overview}
\end{figure}

\section{Introduction}
Mindfulness training, a therapeutic technique that cultivates attention and awareness of the present moment, has gained increasing recognition for its role in improving mental health outcomes ~\cite{lindsay2018acceptance, lindsay2019mindfulness, lindsay2018mindfulness}. Recent studies highlight its potential to mitigate the rising rates of depression, anxiety, and loneliness among young adults in the United States, a population experiencing an unprecedented surge in mental health challenges ~\cite{mann2022loneliness, twenge2021anxiety, twenge2021worldwide}. Epidemiological data reveal that the prevalence of mental illness in this age group is nearly doubling every five to seven years. Compared to standard care and active control groups ~\cite{loucks2021mindfulness, lindsay2018acceptance, mojtabai2016national, lindsay2019mindfulness2, schumer2018brief}, mindfulness training has shown greater efficacy, largely due to higher acceptance and engagement within this demographic ~\cite{cramer2016prevalence}. Clinical trials consistently demonstrate that mindfulness not only reduces stress, loneliness ~\cite{kuyken2015effectiveness, creswell2019mindfulness, creswell2014does}, cortisol levels, and blood pressure reactivity to stress but also significantly enhances daily well-being ~\cite{lindsay2018acceptance, lindsay2019mindfulness, lin2018effects}.

With smartphones now an integral part of daily life for young adults, smartphone-based mindfulness training offers a highly effective, accessible, and practical solution, evidenced by the widespread adoption of mindfulness apps ~\cite{rogers2017internet, plaza2013mindfulness} like Headspace and Calm, which have collectively been downloaded over 180 million times. Randomized controlled trials (RCTs) have demonstrated that these digital interventions can significantly improve both psychological and physical well-being ~\cite{lindsay2018acceptance, lindsay2019mindfulness, lin2018effects}. However, despite their promise, sustained user engagement remains a critical challenge ~\cite{baumel2019objective, nahum2024toward}. For instance, a recent review found that mindfulness meditation apps achieve higher daily usage (median 21.7 minutes) and retention rates compared to other mental health apps ~\cite{baumel2019objective}. Yet, even the retention rate for these apps remains low at just 4.7\%, underscoring a persistent issue with user engagement ~\cite{baumel2019objective}. This highlights the urgent need for more engagement mindfulness applications, as research consistently shows that greater user engagement leads to improved clinical outcomes ~\cite{cox2024mobile, manigault2021examining, creswell2019mindfulness}.

Several decades of research highlight the potential of biosignal feedback to enhance user engagement and outcomes on mobile health platforms ~\cite{kosunen2016relaworld, jarvela2021augmented, huang2018smartphone}. Notable examples include EEG-measured brainwave feedback in the RelaxWorld app ~\cite{kosunen2016relaworld}, EEG and vibration-based haptic feedback in the AttentiveU platform ~\cite{kosmyna2019attentivu}, and respiration biofeedback in gamified platforms ~\cite{shih2019breeze, frey2018breeze}. In the context of mindfulness meditation, respiration tracking emerges as a particularly promising and predictive signal ~\cite{kressbach2018breath, allen2023respiratory}. Although the relationship between mindfulness practice and respiration patterns remains underexplored, preliminary studies suggest distinct alterations in breathing during mindfulness sessions, even when users are not consciously controlling their breath ~\cite{wielgosz2016long, ahani2014quantitative}. This presents an opportunity to provide users with insights into their mindfulness progression and skill development. Despite the growing popularity of commercial mindfulness apps such as Headspace and Calm, which are integrated into wearable devices like Oura rings, Fitbits, and Whoop sensors, few products incorporate real-time biosignal feedback into their training.
While respiration-based biofeedback has been explored in gamified platforms~\cite{shih2019breeze}, it remains absent in both commercial and non-commercial mindfulness meditation apps aimed at improving system usability and user engagement. Existing apps—such as Muse (EEG-based brain signals), Flowly (heart rate and HRV via external sensors), Inner Balance (HRV using HeartMath sensor), and Core Meditation Trainer (ECG)—rely on external biosignal acquisition devices, limiting scalability. Notably, none of these systems incorporate respiration biofeedback, despite its central role in mindfulness practices. Respiration is not only more intuitive and directly connected to meditative techniques like breath awareness, but it can also be captured using minimally obtrusive sensors, making it a practical and scalable feedback modality. Respiration also plays a pivotal role in the physiological manifestation of mindfulness; specifically, slow breathing rates (4–9 bpm) are known to optimize a parasympathetically dominated restful state via vagal stimulation~\cite{gerritsen2018breath, russo2017physiological}, and recent research shows that breathing rhythms modulate brain activity related to emotion regulation and attention~\cite{tort2025global}.

To achieve this goal, this paper addresses three novel research questions: (1) Can respiration biosignals be effectively tracked during digital mindfulness training using the smartphone's built-in sensors? (2) Can these sensor signals predict improvements in mindfulness skills? (3) Does providing respiration biosignal feedback after each session enhance sustained engagement with digital mindfulness training? To answer these questions, we track and visualize respiration biosignal and the change in mindfulness skills as feedback to users into a smartphone application using the phone's on-board accelerometer that delivers a unique mindfulness training curriculum focused on cultivating core skills: concentration— “your ability to focus on what you want to,” sensory clarity— “your ability to track and explore your sensory experience in real-time,” and equanimity— “your ability to allow sensory experiences to come and go,"—based on previous digital mindfulness prototypes ~\cite{lindsay2019mindfulness, lindsay2018mindfulness, lindsay2018acceptance, slutsky2019mindfulness, chi2018effects}. 
As illustrated in Figure~\ref{fig:overview}, our system integrates a feedback module that provides real-time respiration tracking, breathing statistics, and mindfulness progress estimation, allowing users to reflect on their meditation sessions.

To effectively track respiration and mindfulness skills in the app, we address two key technical challenges. First, although respiration tracking has been well studied~\cite{ahmed2021rrmonitor, hernandez2015biowatch, sun2017sleepmonitor, rahman2020instantrr, rahman2021towards, dai2021respwatch, valentine2022smartphone, pechprasarn2013estimation}, existing methods are inadequate for detecting the slow breathing rates (<10 bpm) characteristic of mindfulness practices. Most respiration rate estimation algorithms are developed and tuned for natural breathing rate detection or for monitoring pulmonary patients \cite{rahman2020instantrr, valentine2022smartphone}. To overcome this, we develop an accelerometer-to-respiration rate algorithm capable of accurately tracking both low (4–9 bpm) and regular (10–30 bpm) respiration rates. This algorithm is validated against clinical-grade Hexoskin lifeshirts, demonstrating reliable performance across a wide range of respiration patterns~\cite{jayasekera2021feasibility, khundaqji2020smart}. While respiration biofeedback and smartphone-based tracking have been explored, our work is the first to use respiration as a biomarker for assessing and enhancing mindfulness skills. Moreover, prior smartphone-based methods (e.g., InstantRR~\cite{rahman2020instantrr, valentine2022smartphone}) focus on natural breathing rates (12–20 bpm) and struggle with the slower rates critical to mindfulness practice.

Building on the development of our accurate low respiration tracking, we design a user feedback interface screen that visually represents the change in breaths per minute during each session. It also highlights the total time spent within the optimal breathing zone, which prior research suggests falls between 4 and 9 bpm ~\cite{li2018effects, bernardi2001slow}. Previous studies suggest that this slow breathing zone may optimize a restful parasympathetically-dominated restful state mediated by vagus ~\cite{gerritsen2018breath, russo2017physiological}. In addition to these metrics, participants receive continuous respiratory signal feedback throughout their sessions. While respiration biosignal feedback has been shown to enhance system usability in other contexts ~\cite{shih2019breeze, frey2018breeze}, its application within mindfulness meditation apps has not yet been fully explored. This paper is the first work to integrate respiration feedback and breathing statistics into a mindfulness app.

The second technical challenge is the lack of any quantitative method to measure changes in mindfulness skills—such as concentration, sensory clarity, and equanimity ~\cite{lindsay2018acceptance}—during digital mindfulness training. Current approaches rely on self-report questionnaires ~\cite{mackillop2007further, de2012psychometric}, which are time-consuming, intrusive, and subjective, often leading to lower user engagement. To overcome this limitation, we develop a deep learning (DL) algorithm that uses accelerometer data to estimate changes in mindfulness skills after each guided session. This DL-based solution eliminates the need for self-reporting, providing a scalable and objective way to measure mindfulness progress while minimizing user burden.

In this study, we collect data from 40 participants, completing 261 mindfulness sessions, to evaluate the impact of respiration biosignal feedback on system usability. 
We develop a respiration rate tracking algorithm that not only estimates respiration rate in the range of 4-9 that is crucial for mindfulness meditation, but also accurately tracks natural breathing rate (12-20 bpm). Our evaluation of existing methods shows that the algorithms developed using natural breathing rate fail to detect the low breathing rate. Additionally, we propose the first study to quantify the change in mindfulness skills during meditation using accelerometer data based on respiration changes. We train session-level data to develop the deep learning model to detect the mindfulness change. We integrate the respiration chart, respiration statistics, and mindfulness change in the commercial mindfulness app, Equa, to improve our app's usability, user engagement, and mindfulness experience. We conducted a user study spanning upto 21 days using the session-level feedback. Our results show higher system usability, mindfulness skills change, and user engagement.

The contribution of this paper is as follows:
\begin{itemize}
    \item We develop a respiratory rate detection algorithm that accurately tracks both slow (4–9 BPM) and normal (10–30 BPM) respiration rates, addressing the challenge of monitoring breathing patterns during mindfulness meditation.
    \item We design a deep learning model to estimate mindfulness skill progression—concentration, sensory clarity, and equanimity—using accelerometer-derived respiration data, making it the first approach to quantitatively track mindfulness skill development using physiological signals in digital mindfulness interventions.
    \item We integrate real-time respiration tracking, mindfulness skill assessment, and breathing statistics into a mindfulness app, providing users with personalized feedback to enhance their meditation experience.
    \item We conduct a user study to evaluate the impact of respiration biosignal feedback on system usability, demonstrating its effectiveness in improving engagement and self-awareness. 
\end{itemize}

%% file: IMWUT_25/tex/background.tex
\section{Background and Related Work}

\subsection{Respiratory Biomarker Estimation}
Respiration rate is clinically measured using sensors to detect air pressure near the mouth and nose~\cite{al2011respiration}. Various sensors, e.g., motion~\cite{rahman2020instantrr}, audio~\cite{kumar2021estimating, ahmed2022deep}, and camera~\cite{bae2022prospective}, estimate respiration rate with wearable systems. 

Motion sensors, i.g., inertial measurement unit (IMU), are placed on the chest, head, and wrist to sense breath-related body movement and infer the respiratory parameters from it~\cite{rahman2020instantrr, hernandez2015biowatch, hernandez2015cardiac, sun2017sleepmonitor}. Unlike smartphones, other wearable motion sensors on the chest are not ubiquitous and scalable. Smartphone IMU-based breathing bio-signal, e.g., breathing phase and breathing rate, estimators have shown promising results for pulmonary patients~\cite{chatterjee2020assessing, ahmed2019mlung}. Rahman et al. \cite{rahman2020instantrr} achieves impressive performance using three different algorithms contextually: FFT, peak detection, and Zero cross-detection. However, they only consider short-duration breathing tasks and do not consider low-breathing rate detection. 

Some recent works exploited video-based solutions to estimate breathing rates~\cite{bae2022prospective, sanyal2018algorithms, alnaggar2023video}. Bae et al. ~\cite{bae2022prospective} place a camera at the same height as the participant's face and record the change in motion during inhalation and exhalation. This approach requires adequate lighting conditions and proper placement. Alnaggar et al. \cite{alnaggar2023video} also use video to estimate respiration rate with 1.62 MAE, but the pearson correlation coefficient (PCC) is very low. However, these approaches are unsuitable for passively monitoring mindfulness meditation without imposing additional costs or distracting users from the mindfulness training. Besides, video-based solutions fail to measure low breathing rates from 4-9 bpm.

Smart watches can be a usable solution, as they don’t interrupt the users’ natural interaction with the device. Many previous works used IMU from watches ~\cite{sun2017sleepmonitor} to estimate the breathing rate. While SleepMonitor ~\cite{sun2017sleepmonitor} achieved impressive performance, they evaluated their approach during sleep, which does not incorporate the low breathing rates, which is crucial in mindfulness training. Additionally, most works discard noisy signals by employing signal processing or machine learning techniques, which improves accuracy but also discards breathing sessions. For example, WearBreathing ~\cite{liaqat2019wearbreathing} achieved impressive accuracy (MAE 1.09 BPM) with only 7–16.5\% data retention in a resting position (sitting, standing, and lying) on COPD patients. Such a low retention rate is useless to the user, as they cannot make successful measurements in most attempts. Therefore, we need a more usable solution that estimates the breathing rate with reasonably good accuracy and retention rate. Moreover, many commercial watches do not allow the IMU data from watches with a proper sampling rate, which makes it less desirable for a scalable solution.

Acoustic solutions have also been used for respiratory biosignal estimation ~\cite{breathtrack, breeze, ahmed2023remote}. Recent works on audio-based respiratory signal estimation focus on pulmonary patients~\cite{breathtrack}, guided breathing exercises~\cite{breeze}, or physical exercises~\cite{kumar2021estimating}. Breeze~\cite{breeze} detects breathing phases during controlled breathing exercises, consisting of four seconds of inhalation through the nose, two seconds of exhalation through the mouth, and four seconds of pause. BreathTrack~\cite{breathtrack} develops an acoustic breathing phase detector that is trained using guidance from the IMU collected from the chest. However, BreathTrack focuses on pulmonary patients and aims to differentiate between healthy and clinical participants. The most recent work from Ahmed et al. \cite{ahmed2023remote} achieves 1.72-1.94 mean absolute error (MAE) in estimating respiration rate. However, their work collects data from in-ear microphones and IMU, which are publicly inaccessible and, thus, not a scalable solution. Kumar et al. ~\cite{kumar2021estimating} developed a multi-tasking model to detect respiration rate during exercise from audio signals' Mel-Filter Bank energy (MFB). Breathing audio signals during exercise have some advantages over natural breathing in that they are more audible and have more frequency content. Moreover, none of these works support low respiration rate estimation, which is crucial for predicting mindfulness skills.  

In this paper, we focus on estimating respiration rates using a smartphone accelerometer placed on the chest, offering a ubiquitous and scalable solution with no additional cost. Our goal is to accurately detect low respiration rates (4-9 bpm) alongside natural breathing rates (10-30 bpm), which are crucial for predicting mindfulness skills.

\subsection{Mindfulness Detection}
Over the past two decades, there has been significant growth in scientific research on mindfulness interventions and the development of mindfulness skills \cite{creswell2017mindfulness}. Mindfulness interventions have been shown to increase self-reported mindfulness skills reliably \cite{visted2015impact} and to improve a broad range of health and well-being outcomes in randomized controlled trials \cite{brown2007addressing, creswell2017mindfulness}. There has been considerable debate on the optimal ways to measure mindfulness skill development in this emerging literature. However, in recent mindfulness intervention studies, one approach is to teach people mindfulness skills such as concentration, sensory clarity, and equanimity in a unified mindfulness curriculum. Then, operational definitions of these measures are described to participants in the mindfulness intervention to measure these skills  (i.e., the participants are familiar with them from their mindfulness training programs). Smartphone-based Unified Mindfulness training programs in initial placebo-controlled clinical trials seem compelling \cite{lindsay2018acceptance, lindsay2018mindfulness, lindsay2019mindfulness}. Previous studies utilize trait bases measures of mindfulness skills, such as the trait Mindful Attention Awareness Scale (MAAS)~\cite{mackillop2007further} or Five Facet Mindfulness Questionnaire (FFMQ) ~\cite{de2012psychometric}. These kinds of measurements are not adequate for the large-scale implementation of mindfulness meditation. In this paper, we measure these mindfulness skills sensitively with a single-item measure of each skill before and after each guided mindfulness training session. Thus, in our present work,  we develop an accelerometer signal-based classification model that can predict these changes.

%% file: IMWUT_25/tex/03_method.tex
\begin{figure*}
    \centering
    \includegraphics[width=0.9\textwidth]{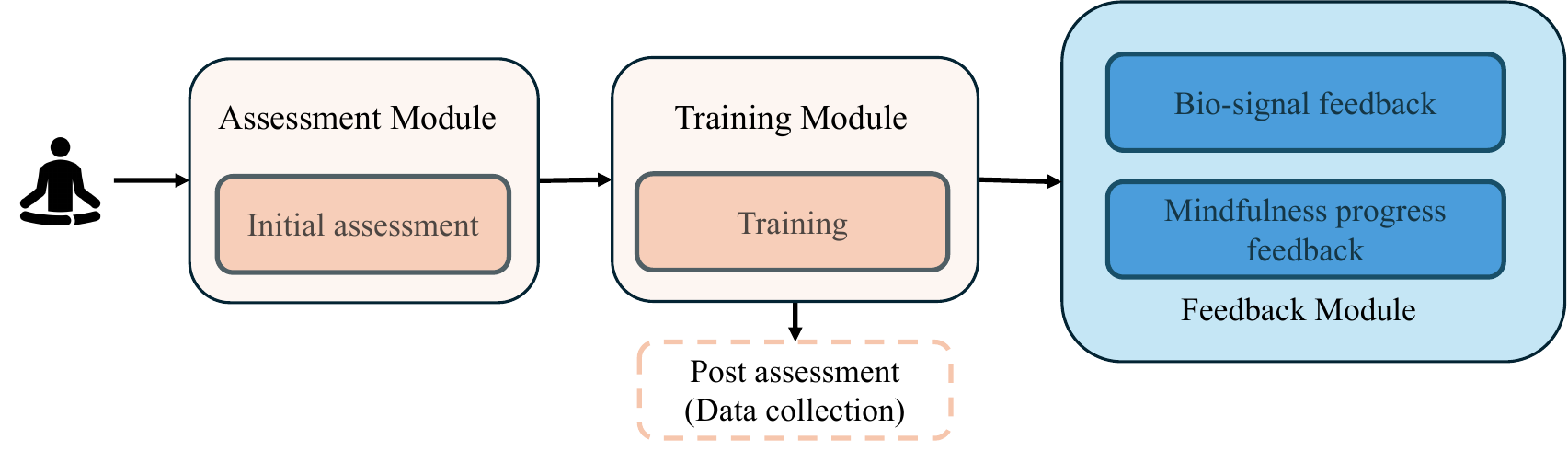}
    \caption{Overview of the smartphone app modules. Post assessments provide ground truth to develop algorithms for the feedback module. The main contribution of this paper is the development, integration, and study of the feedback module highlighted in blue. }
    \label{fig:flow}
\end{figure*}

\section{Overview of the mindfulness training smartphone app}

\begin{figure*}
    \centering    \includegraphics[width=0.8\textwidth]{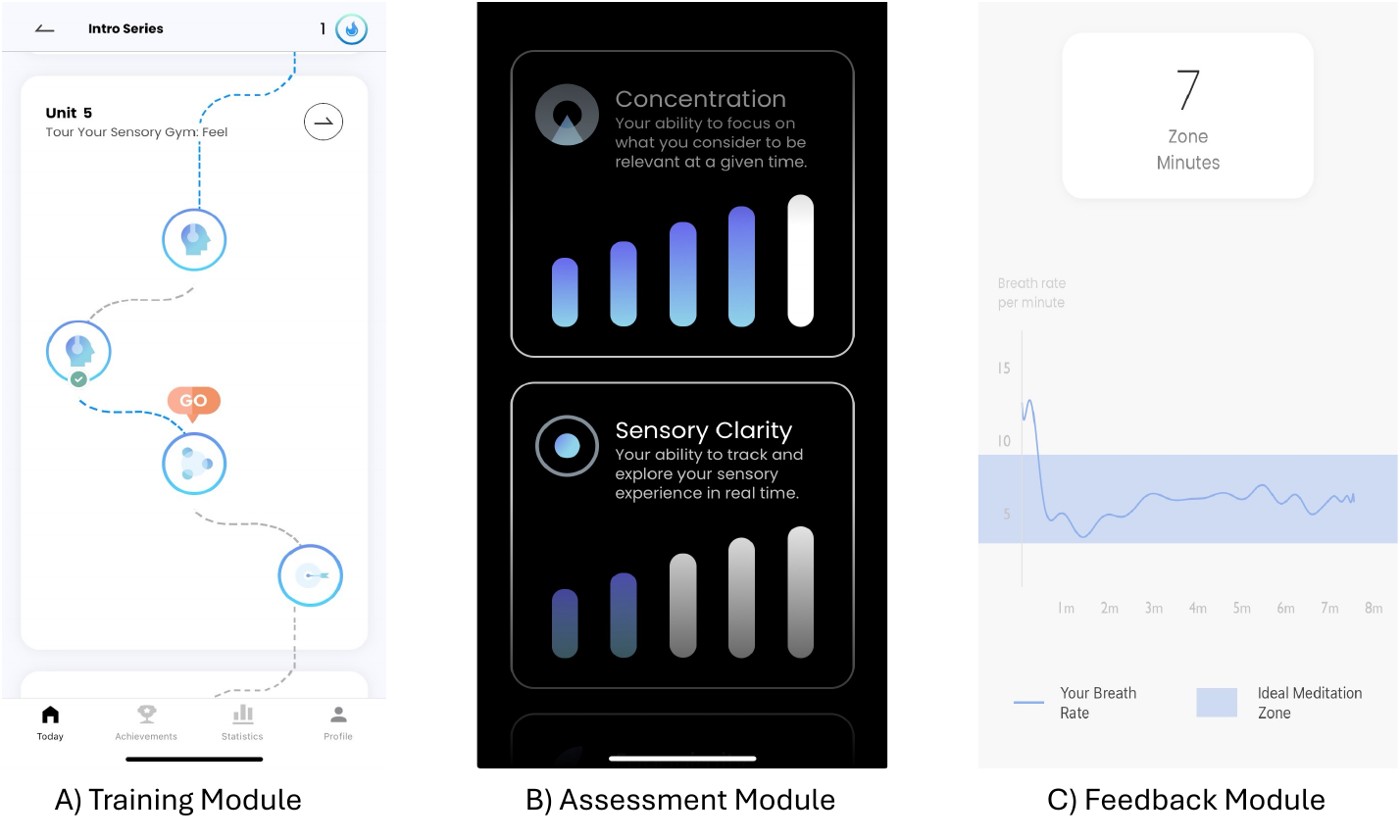}
    \caption{Screenshot of different modules deployed in Equa. The training module consists of necessary instructions for mindfulness meditation training, and the assessment module captures the baseline mindfulness state before the session. Finally, the feedback module provides the users with biosignal feedback such as a respiration chart, respiration statistics, and estimated mindfulness change.}
    \label{fig:app_modules}
\end{figure*}

The mindfulness training app we use in this study is a commercially available app, Equa \cite{equa}, designed to provide evidence-based instruction in training three core mindfulness skills: concentration, sensory clarity, and equanimity. We integrate guided meditation sessions with biosignal feedback and track mindfulness progress in the app. As illustrated in Figures \ref{fig:flow} and \ref{fig:app_modules}, the proposed application delivers a structured curriculum tailored to each user’s mindfulness journey, offering personalized feedback based on their physiological data and session performance. We integrate the feedback module with the app’s existing training and self-assessment modules.  

\parlabel{Training Module}
This module consists of two submodules: (a) training curriculum and (b) meditation training.

a) \textit{Training Curriculum.}
The introductory curriculum consists of 14 interactive and branching lessons. The curriculum has demonstrated an immediate dose-response effect comparable to the Mindfulness-Based Stress Reduction (MBSR) program \cite{kabat1982outpatient, ludwig2008mindfulness}. The curriculum is evidence-based \cite{lindsay2018mindfulness, lindsay2018acceptance, lindsay2019mindfulness, creswell2017mindfulness, lindsay2017mechanisms} and interactive, allowing users to navigate through lessons using screen taps. It also encourages the integration of mindfulness into everyday activities to enhance the long-term success of mindfulness training \cite{lindsay2018mindfulness, manigault2021examining}. By leveraging principles from intelligent tutoring systems \cite{anderson1985intelligent, vanlehn2011relative},  the smartphone app is the first meditation app to offer this type of in-lesson interactivity, in contrast to traditional digital apps that rely primarily on static, one-way guided meditation audio libraries.

(b) \textit{Meditation Training.}
Following the baseline assessment, participants engage in a 10- to 20-minute training session, during which their respiration signals are continuously tracked (see more technical details in section \ref{slow_rr}). 

\parlabel{Self-Assessment Module}
The assessment module aims to collect relevant ground truth to develop algorithms for the feedback module. The app begins with a comprehensive pre-training assessment that evaluates users’ baseline mindfulness skills and includes self-report ratings on the three core mindfulness skills. After each session, users complete the same self-assessment to measure any immediate changes in their mindfulness skills. These self-reported ratings of concentration, sensory clarity, and equanimity serve as ground truth for analyzing the collected respiration data and detecting shifts in mindfulness. While self-reporting is the current way to assess mindfulness skills, and these reports help establish initial benchmarks, a key benefit of mindfulness estimation from accelerometer data is that it minimizes the need for repeated, time-intensive self-assessments after each session—an approach that can reduce user satisfaction in digital mental health apps. By relying on physiological and sensor data to measure mindfulness improvement (see section \ref{mindfulness_dl}), the app aims to streamline the user experience.

\parlabel{Feedback Module}
After each session, participants receive two forms of feedback: respiration biosignal feedback and whether their mindfulness skills improved (or not). 
\begin{itemize}
    \item \textit{Biosignal feedback.} We display a respiration chart from the session, offering users insights into their physiological responses during the guided session.  Our respiration rate tracking algorithm (see section \ref{slow_rr}) is designed to accurately track both slow and natural respiration rates using a smartphone accelerometer while also detecting motion artifacts that may compromise the quality of the data. In such cases, users receive a notification indicating potential data issues, ensuring transparency and trust in the feedback. \newline

    \item \textit{Mindfulness progresses feedback.} The app estimates the user's improvements (or lack thereof) in mindfulness skills, such as concentration, sensory clarity, and equanimity. Our deep learning model allows us to predict the progression of these mindfulness skills based on the smartphone accelerometer data collected during each session. The detailed methodology and performance evaluation of mindfulness change assessment are discussed in the mindfulness progress estimation and algorithmic evaluation sections (see section \ref{mindfulness_dl} and \ref{sec:mindfulness_eval}).
\end{itemize}

\section{Data Collection and Study Protocol}
\label{data collection}
\begin{figure}
    \centering
    \includegraphics[width=0.8\linewidth]{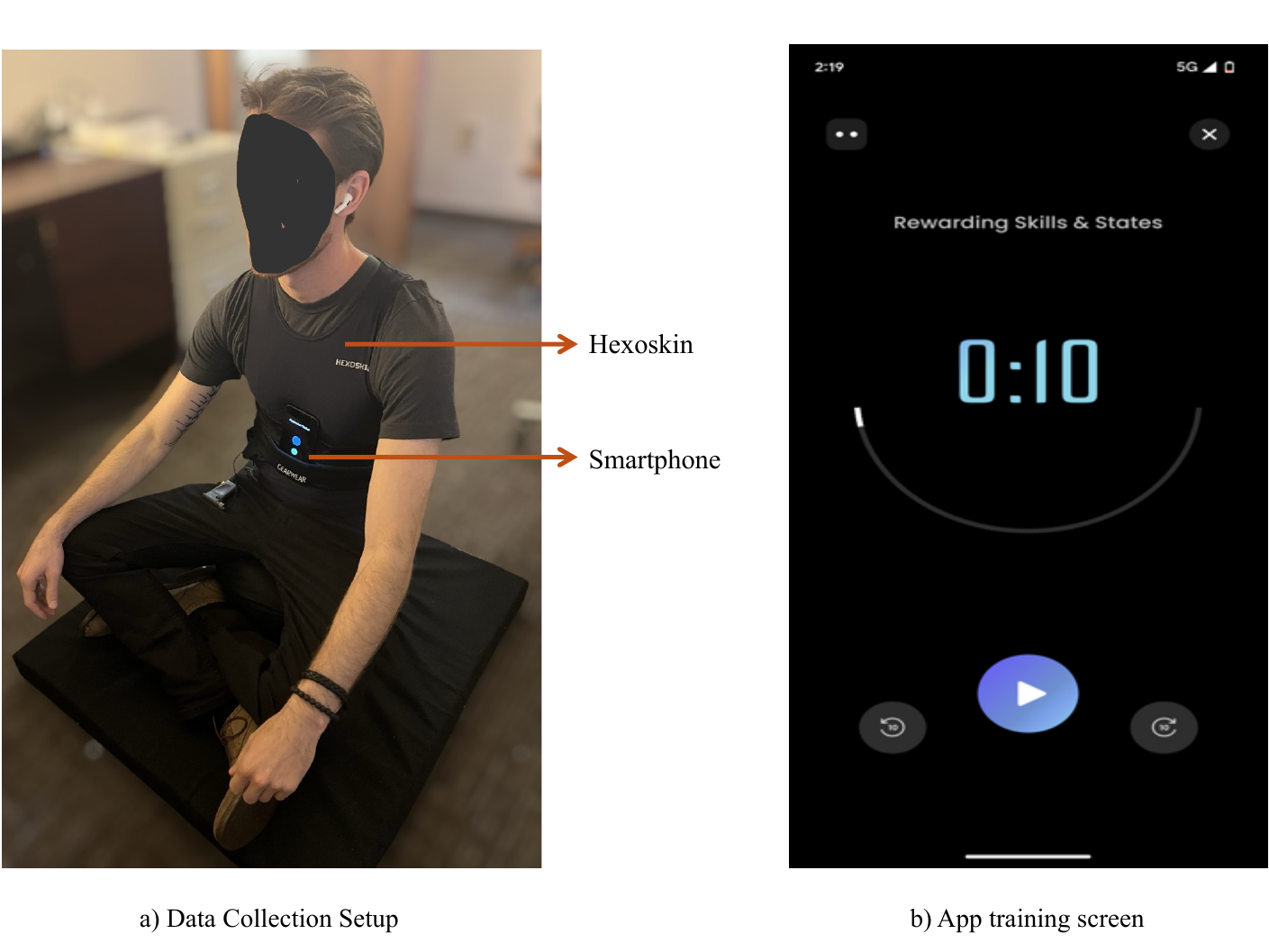}
    \caption{Data collection setup in the lab and smartphone app training screen. Hexoskin is used to capture the ground truth for developing a respiration rate tracking algorithm. The smartphone captures the necessary accelerometer data utilized to develop our proposed algorithms.}
    \label{fig:setup}
\end{figure}

To develop and evaluate our proposed algorithms, we conduct three data collection studies. The first two are conducted in a controlled lab environment, while the remaining two took place in real-world settings with participants using the app. Figure \ref{fig:setup} shows the lab setup and the screen of the app for mindfulness training. 

\newcolumntype{s}{>{\hsize=.2\hsize}X}
\parlabel{Study design and participants}
\begin{table*}[]
\centering
\caption{Participant statistics in different studies. Study 1 and Study 2 are conducted in a controlled environment setting, whereas Study 3 is conducted in the wild. In addition to testing the developed algorithm, study 3 is used for user study.}
\label{tab:study}
\small{
\begin{tabular}{l|l|c|c|c|c|c}
\hline
Study & Environment      & Participants & Sessions & Algorithm Development & Algorithm Testing & User Study \\ \hline
1     & Controlled (Lab) & 8                  & 88 & Yes                        & Yes                 & No                        \\ \hline
2     & Controlled (Lab) & 10                 & 24 & No                         & No               & Yes                        \\ \hline
3     & Natural (home)   & 22                 & 149 & No                         & Yes                & Yes  \\
\hline
\end{tabular}
}
\end{table*}
In total, we collect data from 40 participants, with 18 participating in lab studies and 22 in-the-wild studies. Table \ref{tab:study} shows the participant statistics of all the studies. However, some data samples are not used due to missing ground truth scores or data corruption. The demographic distribution of participants is presented in Table \ref{tab:demographics}. Participants are recruited through community outreach, including online postings, mailing lists, and social media platforms. Eligibility criteria included being fluent in English, being at least 18 years of age, being willing to engage in guided mindfulness meditation sessions, and being willing to wear physiological measuring equipment during meditation training. Initial eligibility screenings were conducted via email, followed by a formal in-person screening at the first session to confirm participation. All participants were first-time users of the app, ensuring consistency in baseline familiarity across study conditions.

\begin{table}[]
\centering
\caption{Breakdown of participants’ demographics in terms of gender, race, and meditation experience in studies 1, 2, and 3. All studies have a balanced distribution of participants.}
\label{tab:demographics}
\begin{tabular}{@{}l|llll@{}}
\toprule
Category              & Group        & Study 1 (\%) & Study 2 (\%) & Study 3 (\%) \\ \midrule
                      & Male         & 50           & 40           & 50           \\
Gender                & Female       & 50           & 40           & 50           \\
                      & Non-binary   & -            & 20           & -            \\ \midrule
                      & Black        & 12           & -            & 4            \\
                      & East Asian   & 12           & 10           & 9            \\
                      & South Asian  & 38           & 20           & 14           \\
Race                  & White        & 25           & 50           & 68           \\
                      & Other        & -            & 10           & -            \\
                      & Mixed        & 13           & -            & 5            \\ \midrule
                      & Not at all   & 12           & 20           & 4            \\
                      & A few times  & 37           & 40           & 23           \\
Meditation Experience & Periodically & 25           & 20           & 41           \\
                      & Frequently   & 13           & 10           & 23           \\
                      & Everyday     & 13           & 10           & 9            \\ \bottomrule
\end{tabular}
\end{table}

a) \textit{Lab study Protocol (Studies 1 and 2)}:
In the lab studies, participants complete a 14-day mindfulness meditation program consisting of 20-minute lessons while their respiration data are collected using a Hexoskin smart shirt and an iPhone X mounted on a chest strap. Participants are randomly assigned into two conditions: a biosignal augmented condition and a control condition. Following the mindfulness lessons, participants in the biosignal augmented condition are exposed to a respiration biosignal feedback chart, which displays the respiration rate trend throughout the lesson. Participants in the control condition are not exposed to this chart. These sessions provided high-fidelity respiration data, which are used as the ground truth for algorithm development. Study 1 included 88 sessions with 8 participants, while Study 2 included 24 sessions with 10 participants. Both studies focused on developing and validating the respiration rate estimation algorithm, testing the accuracy of the mindfulness skill estimation model, and user satisfaction with the app.

b) \textit{In-the-Wild Study Protocol (Study 3)}:
In Study 3, conducted remotely, 22 new participants use the mindfulness meditation app in real-world environments. Similar to the lab study, chest straps and smartphones are used for data collection. This study spans over 21 days. Randomization into the two conditions for study 3 is conducted following the same protocol as in Studies 1 and 2.  Unlike studies 1 and 2, in this study, participants can finish as many sessions as they want and end any session at any time. After 21 days of study, we use the average number of completed sessions and the average number of minutes in meditation to evaluate user engagement. Participants are instructed to complete sessions in quiet, stable environments to minimize movement artifacts. A total of 149 sessions are collected in this study. This study focused on evaluating user satisfaction, user engagement, and the effectiveness of the mindfulness skill estimation algorithm in natural settings.

\parlabel{Physiological and Psychological Data Measures}
Across all studies, chest movements related to breathing are tracked using the smartphone accelerometer. In Studies 1 and 2, physiological data such as respiration rate and motion are also collected using Hexoskin smart shirts, which serve as ground truth for developing the respiration tracking algorithm. Participants also complete self-report assessments of mindfulness skills—concentration, sensory clarity, and equanimity—before and after each session. These self-reported measures provide a baseline for analyzing mindfulness progression and serve as the ground truth for evaluating the deep learning models.

\parlabel{Ethical Considerations}
The study was approved by the Institutional Review Board (IRB) of Carnegie Mellon University. Before data collection began, informed consent was obtained from all participants. All data were anonymized and stored on encrypted servers to ensure participant privacy. Participants were informed about the nature of the data being collected, their right to withdraw from the study at any time, and the potential use of their anonymized physiological data in future research publications.

\section{Slow-Paced Respiration Rate Estimation using Accelerometer for BioSignal Feedback Module}

\label{slow_rr}
\begin{figure*}
    \centering
    \includegraphics[width=0.9\textwidth]{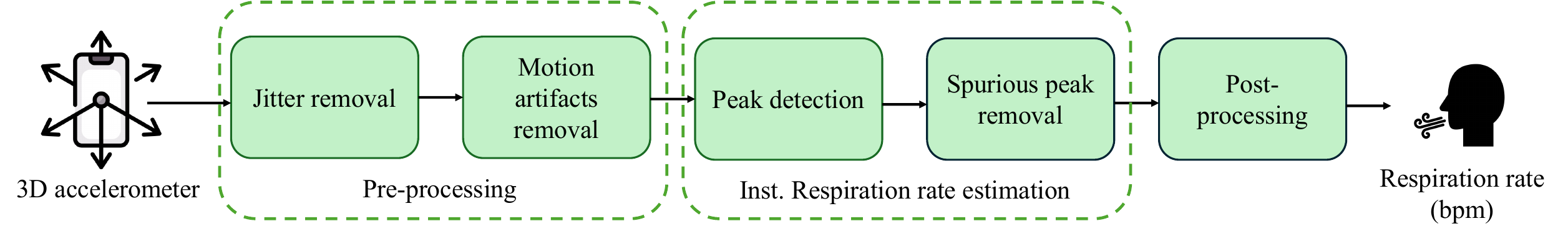}
    \caption{Overview of respiration rate detection algorithm. We pre-process the accelerometer data by removing jitter and motion artifacts before using the proposed peak detection algorithm. Final respiration rates are shown to the user after additional post-processing.}
    \label{fig:rr_flow}
\end{figure*}
In the absence of a reliable algorithm for accurately detecting both slow-paced and regular respiration rates, we develop a novel respiration rate estimation method. Our algorithm leverages accelerometer data, which has been widely utilized in regular or high respiration rate estimation \cite{rahman2020instantrr, rahman2021towards, sun2017sleepmonitor}. However, accelerometer data is often contaminated by two significant noise sources: jitter and motion artifacts, which obscure the respiratory signal—especially for slow respiration rates, where the changes in the accelerometer are smaller and closer to the noise level. To address these challenges, we propose a four-step algorithm for respiration rate extraction. The first two steps involve noise reduction, focusing on removing jitter and motion artifacts. The third step evaluates the quality and reliability of the signal post-noise removal. Finally, the fourth step estimates the respiratory rate from a single sensor stream, providing an accurate measurement of both slow and regular breathing patterns. An overview of the algorithm is shown in Figure \ref{fig:rr_flow}.

\parlabel{Jitter Removal}
Jitter refers to high-frequency noise introduced by inertial sensors, often caused by electrical disturbances such as power supply fluctuations, electromagnetic interference, or errors during analog-to-digital conversion. This noise can interfere with the detection of accurate respiratory signals, potentially leading to incorrect peak detection and distorted frequency content. The effect of jitter is particularly significant during slow breathing, where the subtle changes in acceleration are easily masked by high-frequency noise. To address this, we apply a low-pass Butterworth filter ~\cite{selesnick1998generalized} with a cutoff frequency of 10 Hz. The Butterworth filter is selected for its superior performance in noise reduction \cite{basu2020comparative, gaikwad2014removal, mello2007digital}. 

\parlabel{Motion Artifact Removal}
Motion artifacts, commonly observed in IMU data, are caused by sudden changes in the user’s posture or movement of the phone. These artifacts often produce large amplitude signals that overshadow respiratory signals, making accurate respiration rate extraction challenging. To mitigate this, we apply a local mean removal technique, where the mean of a short window centered around each value is subtracted from the raw sensor measurements. Local mean removal significantly reduces motion artifacts, enhancing the accuracy of respiration rate estimation. For additional denoising, a 13-point moving average filter (i.e., a window of 1.3s) is subsequently applied.

\parlabel{Reliability Assessment} 
Although the previous noise removal steps effectively reduce jitter and motion artifacts, continuous user movement or improper phone placement can still lead to inaccurate respiration rate estimations. To address this, we introduce a signal quality and reliability assessment step, which acts as a safeguard for our respiration rate estimation process. This step is crucial because poor signal quality can distort the entire feedback loop. We assess signal quality by calculating the first and third quartiles, along with the interquartile range (IQR), for each 20s window. If any data value at that window exceeds the third quartile with an addition of 0.8 times the IQR value or falls below the first quartile subtracted by 0.8 times the IQR value, we flag that value as compromised. When compromised values account for more than 25\% of the total data length, the signal is discarded, and a “signal compromised” message is provided to the user. Additionally, we also detect if the phone is on the chest or staying flat on the table. We observe that if the phone is flat on the surface, we get steady and flat accelerometer data with occasional spikes. To identify if the phone is on a flat surface, we take the average of every 30s segment of data and compare it with the average of the whole file. If the difference between the two averages is more than 0.02, we count that as a faulty segment. If the number of faulty segments is lower than 30\% of the total duration of the session, we determine that the data does not have enough variation to be on the chest. We discard the signal with the 'phone is not on chest' message to the user.

\parlabel{Instantaneous Respiration Rate Estimation}
After filtering and signal quality assessment, we apply a peak detection algorithm that identifies all local maxima by comparing each data point with its neighboring values \cite{virtanen2020scipy}. A peak, or local maximum, is defined as any sample whose two immediate neighbors have smaller amplitudes, with each peak corresponding to a breath (inhalation). Our algorithm is optimized to detect respiration rates between 4 and 30 breaths per minute, encompassing both slow and natural respiration rates. To further minimize noise interference, we discard consecutive peaks occurring within 2 seconds and only consider peaks with at least 50\% prominence. Peak prominence is defined as the vertical distance between the peak and its lowest contour line and is used to ensure that only significant chest movements associated with breathing are counted. To compute the respiration rate for each minute, we divide 60 by the duration between two consecutive peaks. We then average the respiration rate over seven cycles to calculate the instantaneous rate, ensuring stability, as respiration rates typically do not fluctuate rapidly unless during intentional breathing exercises. This method reduces errors due to inaccurate peak detection and enhances the robustness of respiration rate estimation.

\section{Estimating Mindfulness Progress via Accelerometer for Mindfulness Progress Assessment Module}
\label{mindfulness_dl}
\begin{figure*}
    \centering
    \includegraphics[width=0.95\textwidth]{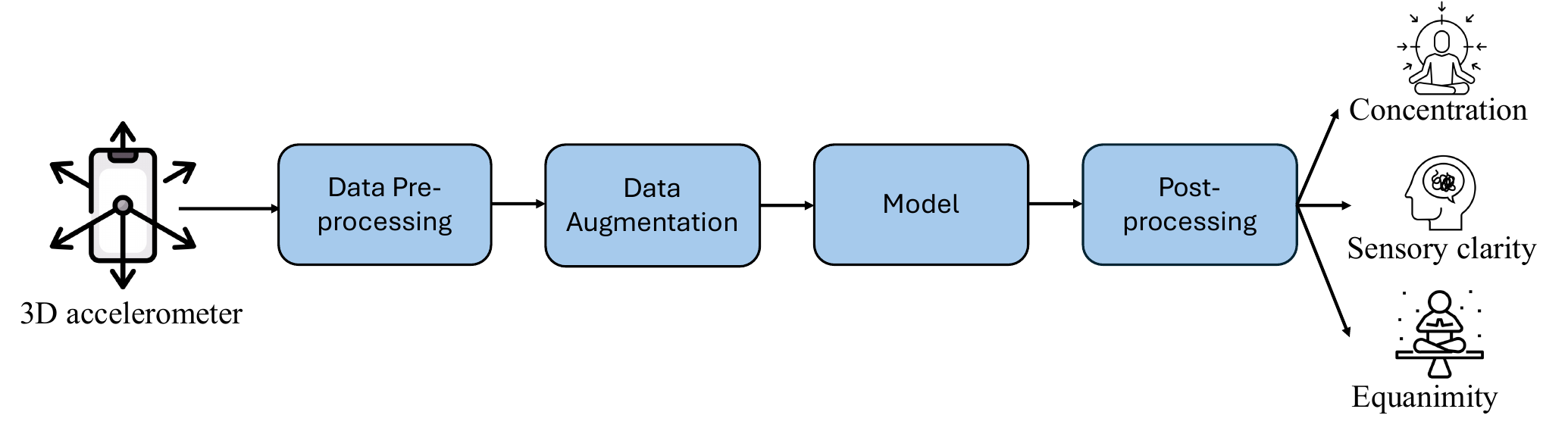}
    \caption{Overview of the algorithm for estimating mindfulness progress via accelerometer data. The data are passed through preprocessing and augmentation before model inference and finally through a post-processing step to estimate changes in concentration, sensory clarity, and equanimity.}
    \label{fig:mind_flow}
\end{figure*}
We propose a deep neural network (DNN) for estimating changes in mindfulness using accelerometer data from a smartphone. This section outlines our data preparation pipeline, the network architecture, training parameters, and post-processing steps involved in our methodology. An overview of the algorithm is shown in Figure \ref{fig:mind_flow}.

\parlabel{Data Pre-processing and Augmentation}
To process the accelerometer data from mindfulness sessions of varying durations, we segment the data into non-overlapping two-minute intervals. This segmentation is based on prior research showing that even brief periods of mindfulness practice, between 60 and 90 seconds, can significantly influence mindfulness levels \cite{crumley2011short}. By using two-minute segments, we ensure that the data is both manageable in size and sufficiently detailed to capture meaningful changes in mindfulness. After segmentation, we obtained 287, 263, and 379 positive samples and 374, 398, and 282 negative samples for concentration, sensory clarity, and equanimity, respectively, revealing a substantial class imbalance between the two classes. To handle this imbalance, we augment the dataset by resampling the class that has a lower amount of data. We use raw accelerometer data rather than the filtered or processed data from the previous section, as the deep learning model is capable of denoising the data itself and capturing relevant information \cite{lecun2015deep, haresamudram2021contrastive}.

\begin{figure*}
    \centering
    \includegraphics[width=\textwidth]{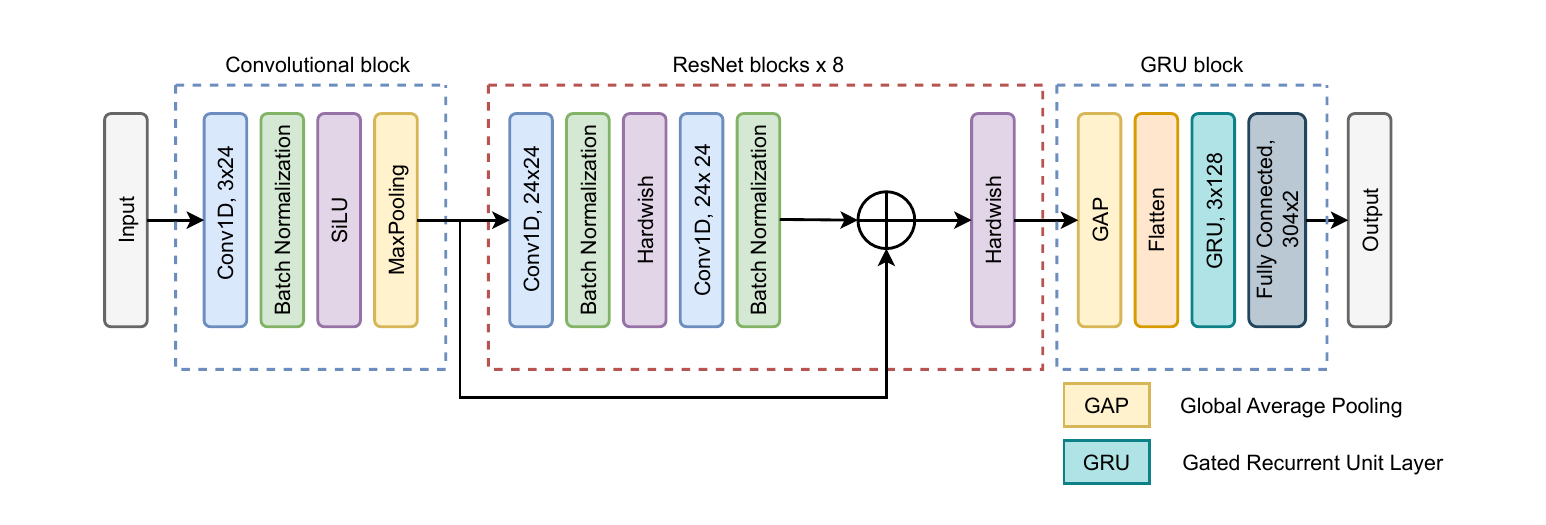}
    \caption{Network architecture for mindfulness skills estimation. The architecture shows a modified version of ResNet \cite{he2016deep} to capture information from accelerometer data with a convolutional block followed by ResNet blocks and a GRU block. GRU block allows the capture of temporal information for the data.}
    \label{fig:nn_architecure}
    
\end{figure*}

\parlabel{Neural Network Architecture}
Figure \ref{fig:nn_architecure} shows the network architecture of the proposed mindfulness progress estimation model that takes the segmented accelerometer data as input and estimates changes in mindfulness skills—concentration, sensory clarity, and equanimity. The architecture combines a 1D ResNet with a Gated Recurrent Unit (GRU) to effectively handle sequential dependencies in time-series signals like accelerometer data.

We modify ResNet architecture (ResNet-1D) \cite{he2016deep} to effectively capture spatial and temporal patterns in the accelerometer data. This ResNet-1D incorporates residual blocks that overcome the vanishing gradient problem and allow the networks to be deeper without loss of performance. Each residual block consists of one-dimensional convolutional layers, batch normalization, and Hardwish activation, which ensures stable and efficient training. The blocks come in two types: identity blocks, where the input is directly passed to the output, and convolutional blocks, which use an additional convolution layer in the shortcut path to adjust the dimensions when needed. The network begins with an initial one-dimensional convolutional layer, followed by eight residual blocks. To reduce dimensionality while preserving key features, global average pooling is applied, and the architecture concludes with a flattened layer for further processing. Following the ResNet-1D layers, a GRU \cite{chung2014empirical} is integrated to capture the temporal dependencies in the IMU data. GRUs, a variant of recurrent neural networks (RNNs) \cite{medsker2001recurrent}, are well-suited for processing sequential data, making them ideal for identifying patterns related to changes in mindfulness skills over time. We employ one GRU layer with 128 hidden neurons. By retaining relevant information from previous time steps, the GRU helps the network effectively recognize shifts in concentration, sensory clarity, and equanimity throughout the mindfulness sessions.

\parlabel{Training and Model Evaluation}
The network is trained using a binary cross-entropy loss function. We use Adam to optimize our model, with a learning rate initially set at 0.0001. We employ a learning rate scheduler to mitigate overfitting by adjusting the learning rate dynamically based on the model’s performance and train the model for 50 epochs. The training is conducted using an NVIDIA RTX 3090Ti GPU, and the best model is saved. The final model is evaluated based on its ability to accurately predict changes in mindfulness skills as compared to self-reported measures, which serve as the ground truth.

\parlabel{Post-Processing and Output Interpretation}
To ensure robust predictions, a majority voting approach was applied during post-processing, where classification outputs from each two-minute segment were aggregated to determine the overall change in mindfulness skills across the entire session.

\section{Integration of feedback Module into smartphone app}
To enable real-time biofeedback and mindfulness skill estimation, we integrated our respiration tracking and mindfulness assessment algorithms into the backend of a commercial mindfulness meditation app (name withheld for anonymity). This integration allows the system to collect sensor data during meditation sessions and provide personalized feedback upon completion.

The app records accelerometer and gyroscope data at \textbf{100} Hz, storing it in Firebase Storage. Structured session metadata and feedback results are maintained in Firebase Firestore (NoSQL). After each session, Google Cloud Functions are triggered via a monitoring service subscribed to the data storage bucket. These functions execute Python scripts that process the sensor data, estimate respiration rates, and compute mindfulness scores. The processing typically completes within \textbf{0.2} seconds, and results are written back to Firestore via a RESTful API, enabling the app to update feedback visualizations (e.g., trend graphs, score tables) for the user. Feedback is provided post-session rather than in real-time to avoid interrupting the meditation experience and to ensure robust data processing.

The mobile app is cross-platform (iOS and Android), with the frontend dynamically retrieving feedback data from Firestore APIs. The backend architecture scales horizontally using Google Cloud’s serverless infrastructure to support concurrent users without significant latency. To maintain data security and privacy, all sensor data is anonymized and transmitted using encrypted channels, following standard company privacy policies. Sessions with insufficient sensor data quality are flagged, and feedback is withheld to maintain accuracy.

The architecture and integration workflow are modular to support integration with other mindfulness or biofeedback applications. Backend algorithms and deployment pipelines are maintained through version control systems, supporting continuous deployment and future updates. The design also allows for easy integration of additional sensors or algorithms in future iterations.

%% file: IMWUT_25/tex/05_result.tex
\section{Results}
In this section, we present the technical evaluation of our proposed algorithms. To assess the effectiveness of our proposed algorithms\footnote{Codes and dataset can be found at \href{https://bashlab.github.io/meditite_project/}{https://bashlab.github.io/meditite\_project/}}, we conduct a comprehensive evaluation using the data collected from our studies. Specifically, we measure the accuracy of the respiration rate estimation algorithm and evaluate the mindfulness skill change detection algorithm’s ability to identify shifts in mindfulness skills using accelerometer data, demonstrating their effectiveness using data from both controlled laboratory settings and real-world environments. 

\subsection{Respiration Rate Estimation Algorithm}
We assess the performance of our respiration rate estimation algorithm using data collected from study 1. We first compare the performance of our proposed algorithm against the state-of-the-art (SOTA) respiration detection algorithms using IMU. Then, we dive deeper into the performance analysis of our algorithm by analyzing it against the ground truth itself.

\parlabel{Evaluation Metrics}
We employ two widely recognized metrics, Mean Absolute Error (MAE) and Pearson Correlation Coefficient (PCC), to evaluate the performance of our algorithm.
\begin{itemize}
    \item MAE is a common metric used in respiration rate estimation studies \cite{liaqat2019wearbreathing, rahman2020instantrr, progga2023meteorological}. It measures the average absolute value difference between the estimated respiration rates and the ground truth values. MAE is defined as:
        \begin{equation}
            MAE = \sum^{n}_{i=1}\frac{|BR_{gt_i} - BR_{est_i}|}{n}
        \end{equation}
    Here, n is the number of samples, $BR_{est}$ is the estimated respiration rate, and $BR_{gt}$ is the ground truth respiration rate. A lower MAE indicates higher accuracy in estimation. In this study, we calculate the MAE by comparing the algorithm’s estimated respiration rates to the ground truth data obtained from the Hexoskin smart shirts.
    
    \item PCC measures the linear correlation between $BR_{gt}$ and $BR_{est}$ \cite{sedgwick2012pearson}. PCC is an important metric for understanding the algorithm’s performance to identify respiration rates for long sessions. PCC provides insight into how well the predictions follow the breathing patterns. PCC is defined as:
    \begin{equation}
        PCC = \frac{cov(BR_{gt},BR_{est})}{std(BR_{gt})std(BR_{est})}
    \end{equation}

    Here, $std$ and $cov$ refer to standard deviation and covariance, respectively. The PCC value ranges between $-1$ and $+1$ (1) A PCC closer to 1 indicates a strong positive linear relationship, (2) A PCC near 0 suggests no linear correlation, and (3) A PCC closer to $-1$ indicates a strong negative linear relationship. A higher PCC (closer to 1) signifies that the estimated respiration rates closely follow the patterns of the ground truth, demonstrating the algorithm’s effectiveness in tracking respiration trends over time.
\end{itemize}

\begin{figure*}
    \centering    \includegraphics[width=0.9\textwidth]{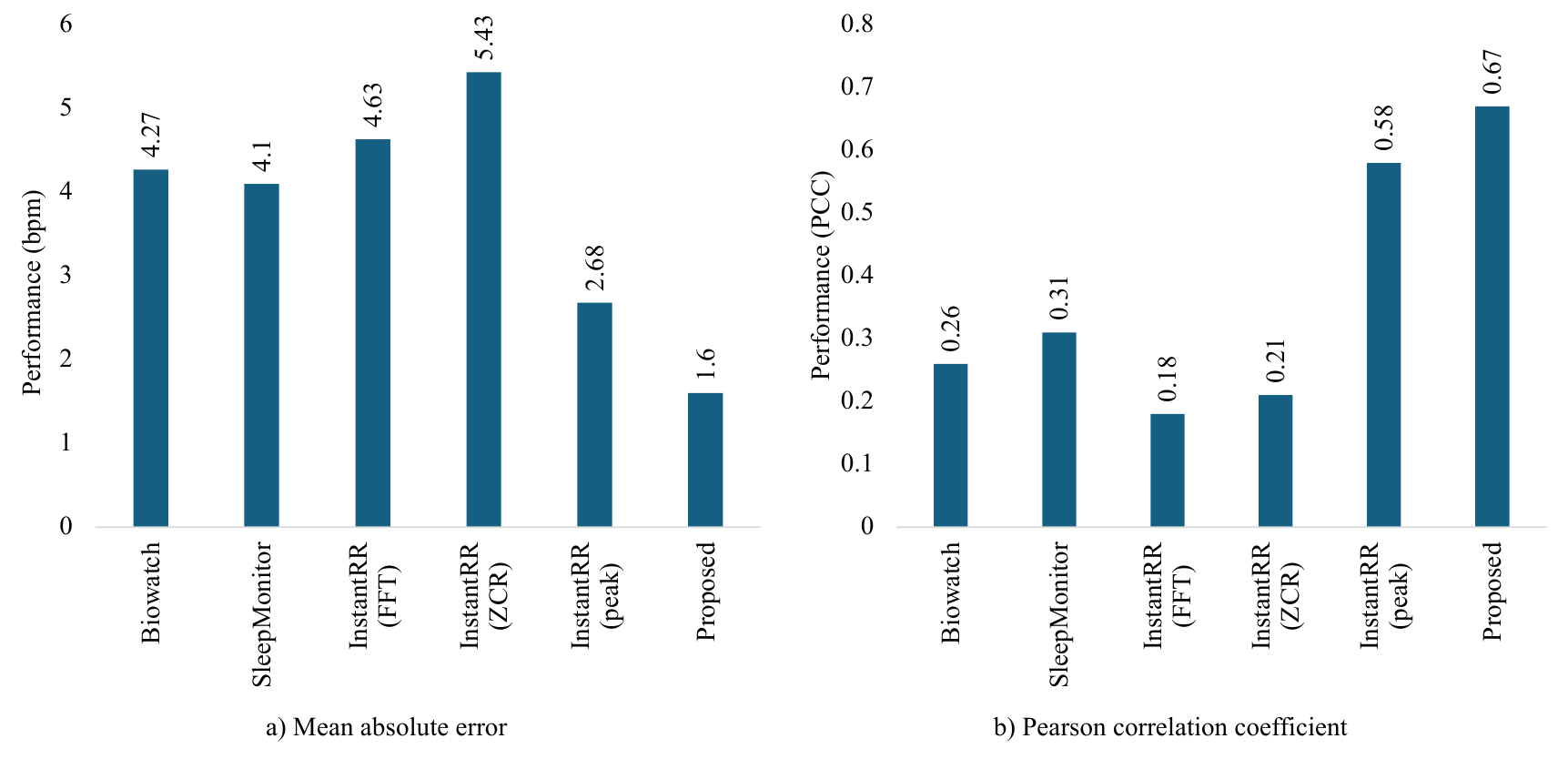}
    \caption{Performance comparison between the proposed algorithm and baseline algorithms. The proposed algorithm demonstrates less MAE and high PCC compared to existing SOTA methods.}
    \label{fig:mae_pcc_rr}
\end{figure*}

\begin{figure}
    \centering
    \includegraphics[width=0.9\linewidth]{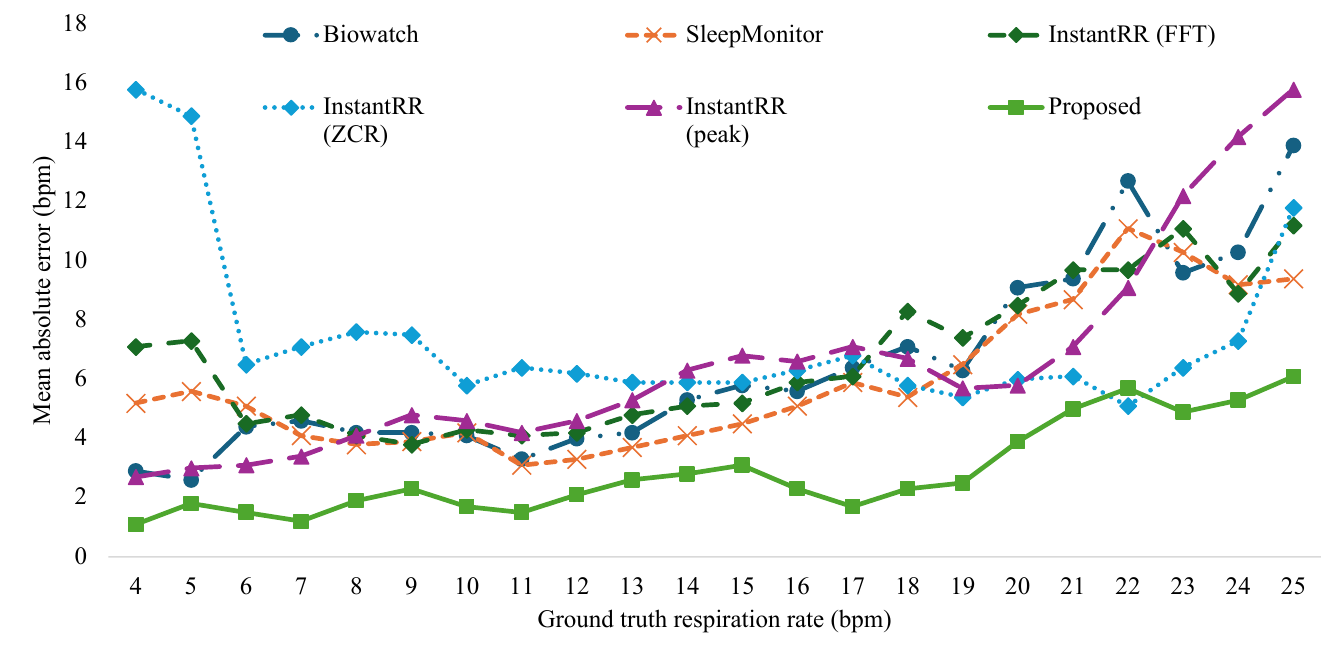}
    \caption{The mean absolute error for different respiration rates shows that the proposed algorithm performs consistently over a wide range of respiration rates, including low respiration rate zones with a lower MAE compared to the SOTA methods.}
    \label{fig:mae_by_bpm}
\end{figure}

\parlabel{Comparison against SOTA IMU To Respiration Rate Detection Algorithms}
We evaluate the performance of our respiration rate estimation algorithm against three state-of-the-art accelerometer-based methods: (1) BioWatch \cite{hernandez2015biowatch}, (2) SleepMonitor \cite{sun2017sleepmonitor}, and (3) InstantRR \cite{rahman2020instantrr}. Both BioWatch and SleepMonitor employ Fast Fourier Transform (FFT)-based algorithms on IMU data collected from smartwatches. InstantRR utilizes a combination of FFT-based algorithms, peak-detection algorithms, and Zero-Crossing Rate (ZCR) algorithms for different body positions, each with specific data processing techniques. As illustrated in Figure \ref{fig:mae_pcc_rr}, our method outperforms BioWatch, SleepMonitor, and all three algorithmic variations of InstantRR. Specifically, our algorithm achieves 13-73\% higher PCC and reduces MAE by 40-70\% across various respiration rates. This significant improvement is attributed to our algorithm’s robust performance in accurately estimating both slow and regular respiration rates. 

Figure \ref{fig:mae_by_bpm} illustrates the average performance of all algorithms across different respiration rates observed in study 1, aggregated over all participants. The results demonstrate that our algorithm consistently maintains a low MAE across the entire range of respiration rates, effectively detecting both slow and natural breathing patterns. Notably, our algorithm achieves less than a 2 BPM error up to 19 BPM. In contrast, the compared methods perform adequately within the 8–15 BPM range, but exhibit decreased accuracy at both low respiration rates—which are prevalent during mindfulness practices—and very high respiration rates. Their reliance on FFT and ZCR techniques, which have insufficient frequency resolution and increased susceptibility to noise, makes them less effective for slow-paced respiration. In contrast, our algorithm effectively captures slow breathing patterns through tailored noise reduction techniques and adaptive peak detection mechanisms. By addressing the challenges of low-amplitude signals and minimizing the impact of motion artifacts, our method provides more reliable respiration rate estimations in contexts where slow breathing is prevalent, such as mindfulness training. 

\begin{figure}
    \centering
    \includegraphics[width=0.8\textwidth]{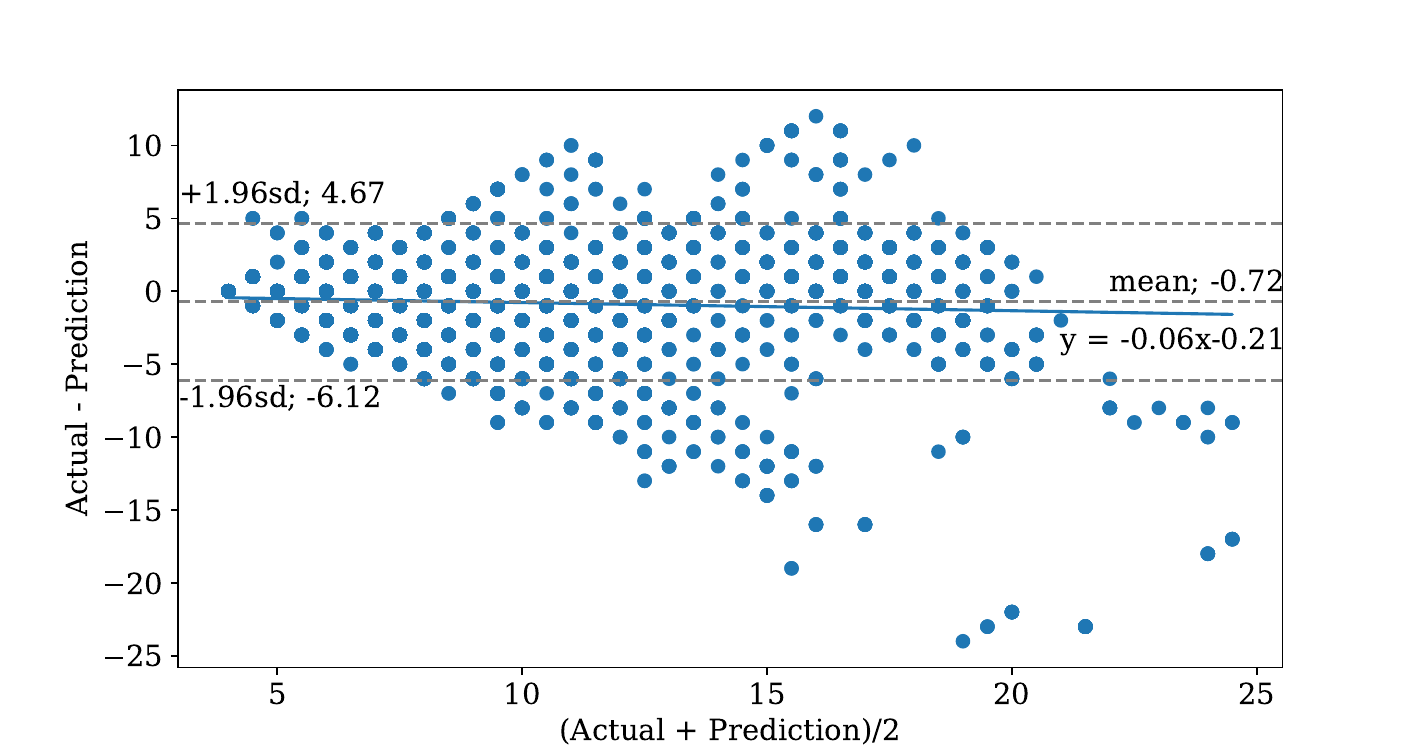}
    \caption{Bland-Altman plot for respiration rate estimation using the proposed algorithm shows that most of the data are within the limits of agreement, indicating the effectiveness of our proposed algorithm.}
    \label{fig:bland_altman}
\end{figure}

\parlabel{Agreement Analysis Using Bland-Altman Plot}
To further assess the agreement between our estimated respiration rates and the ground truth, we employ a Bland-Altman plot \cite{bland1986statistical}. The Bland-Altman plot is a method used to analyze the agreement between two quantitative measurement techniques by plotting the difference between the methods against their mean. Figure \ref{fig:bland_altman} presents the Bland-Altman plot for our data. We plot a random sample of data points for clarity; however, the entire dataset is used to calculate mean differences and limits of agreement. Most of our algorithm’s estimations are within the limits of agreement. Combining these findings with the high PCC reported earlier, we conclude that there is substantial agreement between our estimated respiration rates and the ground truth data from Hexoskin. This validates the accuracy and reliability of our algorithm in estimating respiration rates across various breathing patterns.

\parlabel{Ablation Study and Parameter Tuning}
This section examines the impact of various parameters on the algorithm design, providing rational explanations for our design decisions.

\begin{figure}[!htb]
\centering
    \subfloat[local mean removal\label{local mean}]{\includegraphics[width=.48\linewidth]{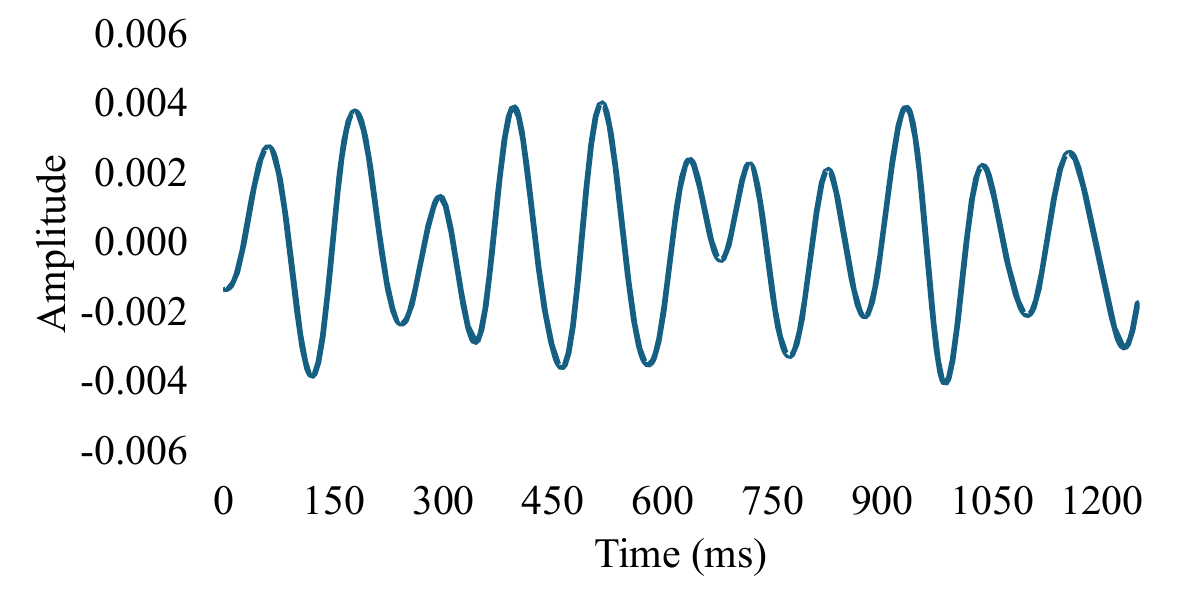}}
    \subfloat[global mean removal \label{window}]{\includegraphics[width=.48\linewidth]{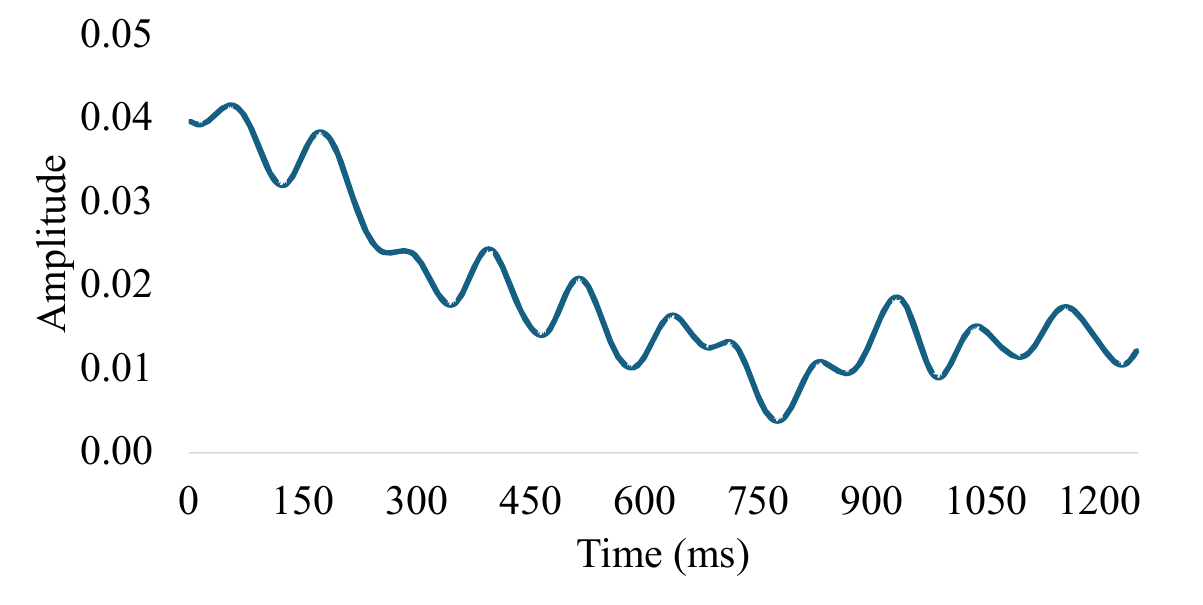}}
    \caption{Effect of applying a) local mean removal, b) global mean removal techniques on data preprocessing. Using the local mean removal can remove the motion artifacts more efficiently than using the global mean removal technique.
    }
    \label{mean_removal}
\end{figure}

\noindent\textit{Effect of Local Mean Removal.}
Figure \ref{mean_removal} illustrates the rationale and effectiveness of applying local mean removal compared to global mean removal for motion artifact suppression. Global mean removal assumes stationarity over the entire signal, which is often invalid during respiration monitoring, especially in the presence of motion artifacts or shifts in baseline. In contrast, local mean removal dynamically adapts to temporal fluctuations, allowing for better isolation of the respiration component from transient noise and motion disturbances. Local mean removal significantly reduces respiration rate estimation error, validating its role in enhancing signal quality for reliable biosignal tracking in real-world scenarios.

\begin{figure}[!htb]
\centering
    \subfloat[Prominence\label{prominence}]{\includegraphics[width=.48\linewidth]{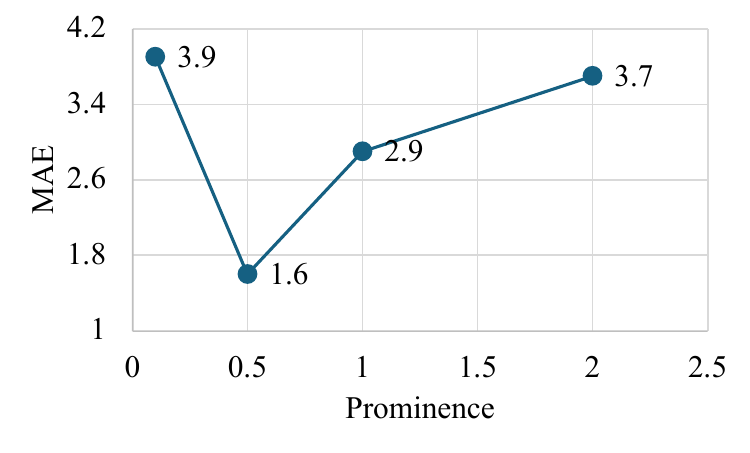}}
    \subfloat[Averaging window \label{window}]{\includegraphics[width=.48\linewidth]{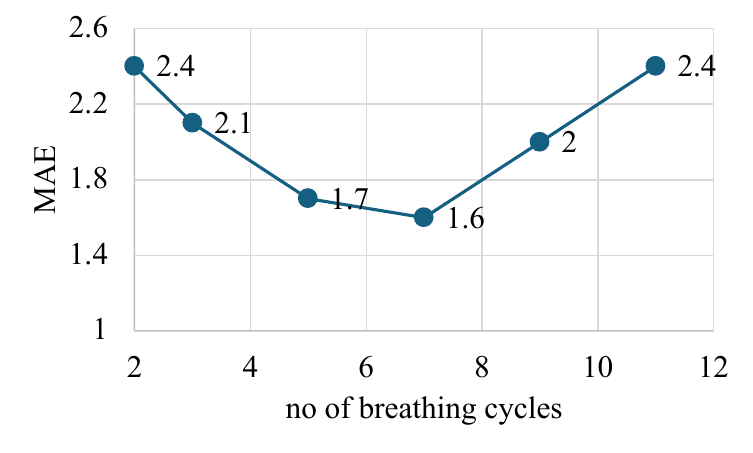}}
    \caption{Effect of a) Prominence and b) Averaging window (breathing cycles) on the performance of the respiration rate algorithm allows us to select the optimized parameters to estimate the respiration rate with high accuracy.
    }
    \label{prominence_window}
\end{figure}

\noindent\textit{Effect of Prominence.}
The prominence parameter plays a crucial role in peak detection for accurate respiration rate estimation. Figure \ref{prominence} illustrates how varying prominence values affect algorithm performance. Setting the prominence too high suppresses smaller, yet valid, respiration peaks, leading to missed detections and degraded performance. Conversely, a low prominence threshold allows excessively small-amplitude peaks, often caused by noise or motion artifacts, resulting in false detections. Through systematic fine-tuning, we identified an optimal prominence value of 50\%, which balances sensitivity and specificity, achieving the best respiration rate estimation accuracy in our dataset.

\noindent\textit{Effect of Averaging Window (Breathing Cycles).}
Figure \ref{window} illustrates how the number of breathing cycles used for averaging impacts respiration rate estimation performance. Using fewer cycles enables more immediate, responsive estimates, but increases susceptibility to errors from false peak detections. In contrast, averaging over a larger number of cycles smooths out these fluctuations, improving accuracy but introducing latency in the feedback. To balance responsiveness and stability, we empirically determined that averaging over seven breathing cycles provides optimal performance with minimal delay, ensuring reliable respiration rate feedback without compromising timeliness.

\noindent\textit{Effect of Multiplier on Data Reliability Assessment.}
To assess the reliability of sensor data, we compute the inter-quartile range (IQR) within each window and apply a multiplier to define the acceptance threshold. Specifically, we use a multiplier of 0.8, selected empirically to balance data retention and algorithm performance.

\begin{figure*}[!htb]
\begin{minipage}{0.33\textwidth}
\centering  
\includegraphics[width=\textwidth]{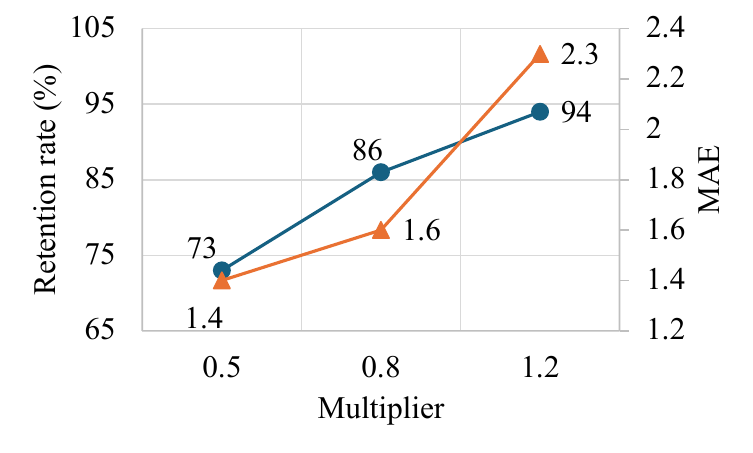}
    \caption{Effect of multiplier on data retention rate and performance of algorithm. }
    \label{fig:multiplier}
\end{minipage}
\begin{minipage}{0.33\textwidth}
\centering  
\includegraphics[width=\textwidth]{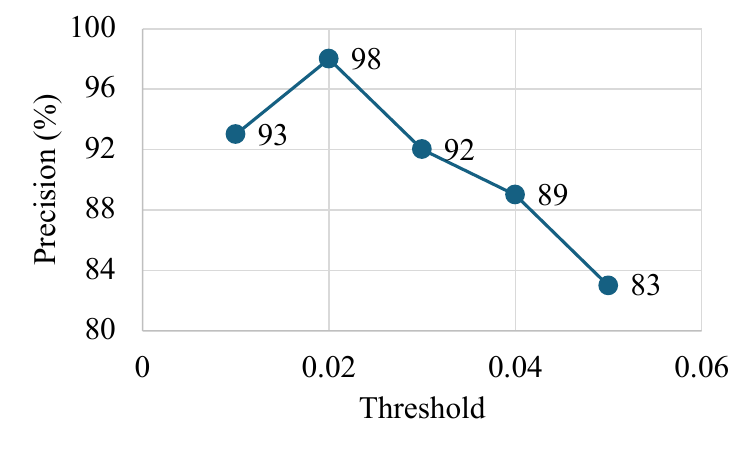}
    \caption{Effect of threshold on the performance of detecting flat surface.}
    \label{fig:flat_thres}
\end{minipage}
\begin{minipage}{0.32\textwidth}
    \centering
    \includegraphics[width=\textwidth]{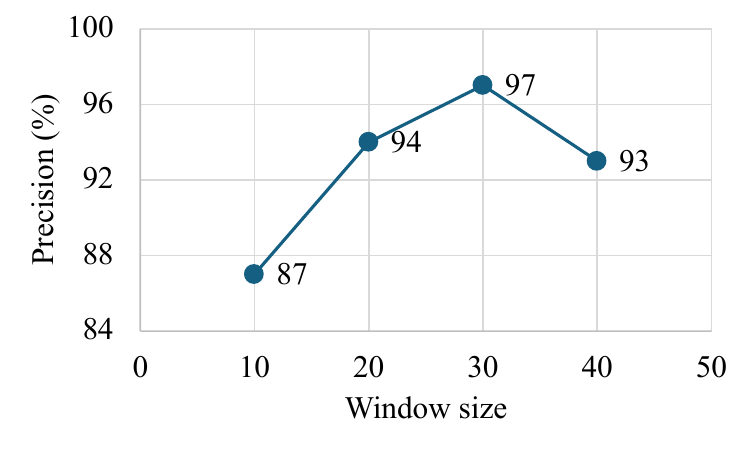}
    \caption{Effect of window-size on the performance of detecting flat surface.}
    \label{fig:flat_window}
\end{minipage}

\end{figure*}

Figure \ref{fig:multiplier} illustrates how varying the multiplier affects the retention rate of data windows. A smaller multiplier imposes a stricter reliability threshold, effectively filtering out noisy or unreliable data. However, this significantly reduces the retention rate, which is impractical for real-world applications as it limits the amount of usable data, potentially degrading system usability and user engagement.

Conversely, a larger multiplier relaxes the threshold, increasing the retention rate but allowing more noisy data to pass through. This can lead to a drop in the performance of both the respiration rate estimation and mindfulness skill assessment algorithms, negatively impacting user satisfaction due to less reliable feedback.

Through empirical evaluation, we identified 0.8 as the optimal multiplier, ensuring a reasonable retention rate without compromising the accuracy of feedback provided to users.

\noindent\textit{Parameter Sensitivity Analysis in Flat Surface Detection.}
Flat surface detection acts as an additional data reliability checkpoint, ensuring that sensor data is processed only when the phone is appropriately placed on the user’s chest, not on unintended surfaces. This mechanism improves system usability by preventing unreliable data from influencing feedback.
To develop and fine-tune the flat surface detection algorithm, we augmented our dataset with 20 sessions collected while the phone was placed on various flat surfaces (e.g., table, floor, book). We used 15 of these sessions along with 15 valid meditation sessions from Study 1 for training, and tested on the remaining 5 flat-surface sessions and 15 additional meditation sessions.

Figure \ref{fig:flat_thres}, and \ref{fig:flat_window} present the sensitivity analysis of the detection threshold and window size parameters, focusing on precision performance. Precision is prioritized to minimize false positives, ensuring that valid meditation sessions are not mistakenly classified as flat surface instances, which would otherwise result in data loss and reduced user satisfaction.
Through this analysis, we identified an optimal balance between detection sensitivity and data retention, ensuring reliable flat surface identification without compromising the inclusion of valid sessions.

\parlabel{Discussion}
Previous works on breathing rate detection mainly target natural or patient breathing, where chest movements are more prominent, making detection easier. Natural breathing rates (12–20 bpm \cite{nicolo2020importance}) and pulmonary conditions \cite{rahman2020instantrr} often produce strong signals, allowing frequent peak detection in accelerometer data. However, focusing solely on high breathing rates risks losing low-frequency inhale/exhale peaks during filtering. As prior algorithms were optimized for natural rates, their performance degrades at low breathing rates. In contrast, our algorithm is specifically designed for low breathing rates, with parameters fine-tuned to capture peaks during both low and natural breathing. Additionally, careful selection of prominence to capture small amplitude peaks and use of an averaging window to post-process the respiration rate ensures high performance in detecting low breathing rates. When a user has a low breathing rate, actual peaks occur after a long duration (15s for 3 bpm). Any instantaneous peaks during this period will risk the accuracy of the algorithms. The use of an averaging window helps us to mitigate this erroneous peak detection and maintain high performance in a low-breathing-rate zone.

\parlabel{Implications and Conclusion}
Overall, these results confirm that our algorithm not only maintains low MAE across different respiration rates but also shows strong agreement with established measurement techniques. This underscores its effectiveness in providing accurate respiration rate estimations, which is critical for applications like mindfulness training, where both slow
and natural breathing patterns are prevalent. 

\subsection{Mindfulness Progress Estimation Algorithm}
\label{sec:mindfulness_eval}
We assess the proposed algorithm's performance by comparing its effectiveness against other network architectures and against other data modalities.

\parlabel{Evaluation Metrics}
We evaluate the proposed algorithm using three popular metrics: precision, recall, and F1 score \cite{khan2024sound, wardhani2019cross, khan2024infantmotion2vec, biswas2025quads}.
\begin{itemize}
    \item Precision measures how many of the predicted positives are truly positive. Precision is defined as:
    \begin{equation}
        Precision = \frac{TP}{TP+FP}
    \end{equation}
    Here, TP and FP stand for true positive and false positive, respectively.
    \item Recall measures how many of the actual positives the model correctly identified. Recall is defined as:
    \begin{equation}
        Recall = \frac{TP}{TP+FN}
    \end{equation}
    Here, FN refers to false negatives.
    \item F1--score is the harmonic mean of precision and recall, balancing both to give a single performance measure, particularly useful in cases of imbalanced data. F1 score is calculated as:
    \begin{equation}
        F1-score = 2 \times \frac{Precision \times Recall}{Precision+Recall}
    \end{equation}
\end{itemize}

\begin{figure*}
    \centering
    \includegraphics[width=\textwidth]{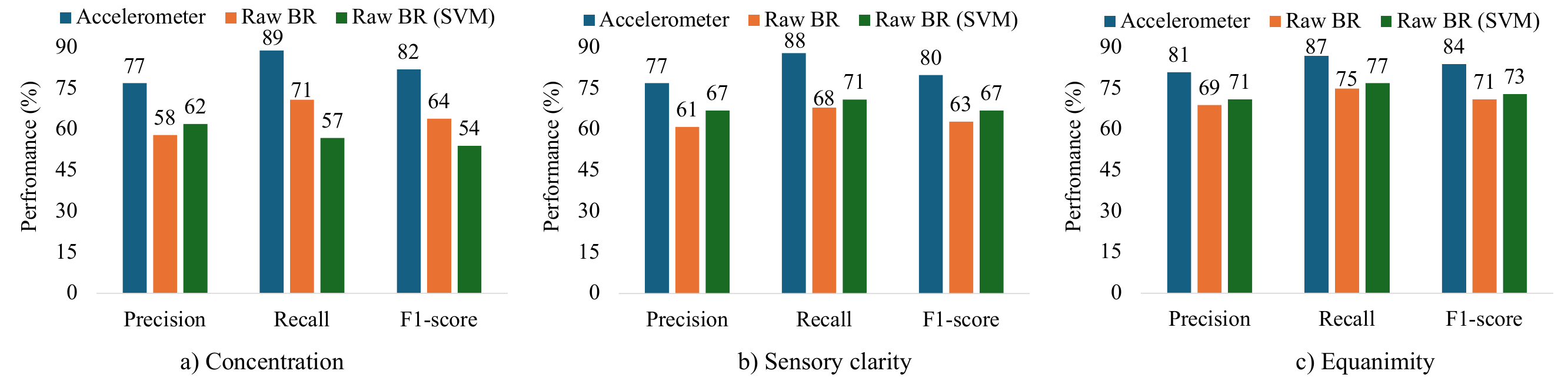}
    \caption{Performance comparison of smartphone accelerometer with estimated respiration for mindfulness change estimation. Using accelerometer data over raw respiration rate performs better in all three mindfulness change predictions.}
    \label{fig:mode_all}
\end{figure*}
\begin{figure*}
    \centering
    \includegraphics[width=\textwidth]{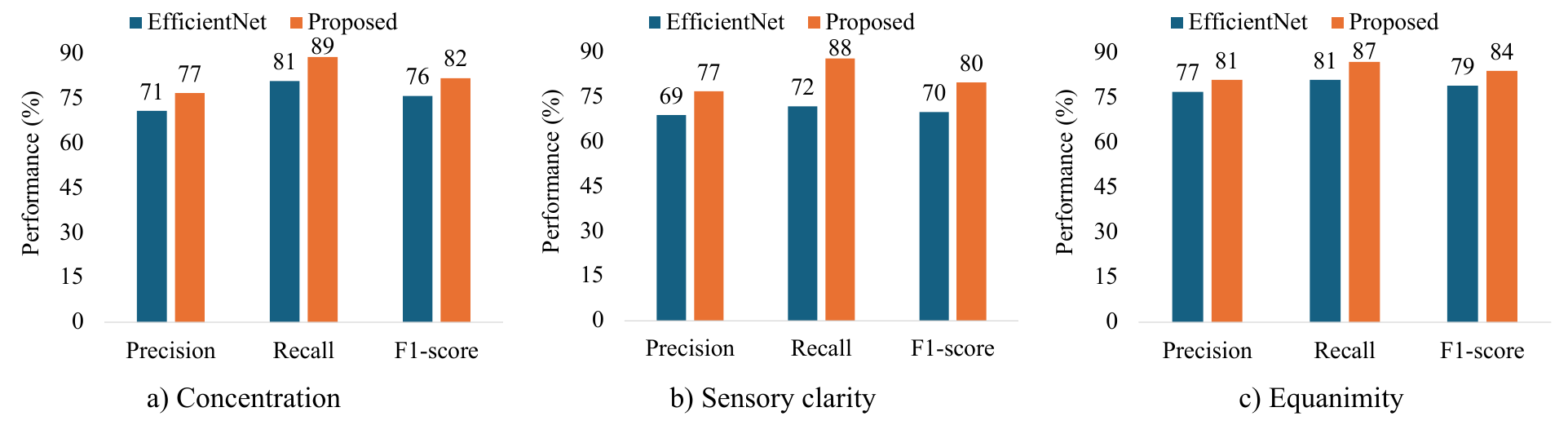}
    \caption{Performance comparison of various models for estimating change in mindfulness skills. Our proposed model demonstrated superior performance with 80\%-84\% F1 score, outperforming others and highlighting its effectiveness in capturing mindfulness skill progression.}
    \label{fig:net_all}
\end{figure*}

\parlabel{Comparison with Ground Truth}
Figure \ref{fig:mode_all} illustrates the results of our proposed model against the self-reported ground truth. Our model achieves F1-Scores of 82\% for concentration, 80\% for sensory clarity, and 84\% for equanimity. We observe that the change in  equanimity is predicted more accurately by the model than the other two skills. This higher accuracy in predicting equanimity may be attributed to the distinct physiological and behavioral markers associated with this skill during mindfulness practice.  Equanimity involves a balanced state of emotional regulation and reduced reactivity to stressors, which may manifest as steadier breathing patterns and minimal physical movement—signals that are effectively captured by accelerometer data. The model’s ability to detect these subtle cues underscores the effectiveness of our model to monitor mindfulness skill progression.

\parlabel{Comparison of Data Modalities}
We compare our DL model for estimating mindfulness skill improvements using two different data sources: raw accelerometer data from smartphones and true respiration rate data from Hexoskin smart shirts. The inclusion of Hexoskin data serves a crucial purpose—we want to eliminate any potential errors arising from respiration rate estimation using raw accelerometer data. By using accurate respiration measurements from the Hexoskin Smart shirts, we aim to understand the true effectiveness of our algorithm without the confounding effects of estimation inaccuracies. The models are trained on data from study 1, and inference is conducted on an independent set of data from study 3 (see table \ref{tab:study}). For each modality, we explore various network architectures and report the best results. We choose to compare our accelerometer-based model with models using true respiration rates because respiration rate is a well-established physiological indicator associated with mindfulness and relaxation states. By including respiration rate as a baseline modality, we aim to assess whether accelerometer data can provide additional predictive power over this traditional measure. Comparing the two modalities allows us to determine the extent to which accelerometer data captures relevant information for mindfulness skill estimation beyond what is offered by respiration rates alone.

For the accelerometer modality, we employ our proposed ResNet-1D architecture combined with a GRU. The raw accelerometer data provides a rich and dense dataset with ample data points per segment, making it well-suited for this DL approach. In contrast, the true respiration rate data from the Hexoskin smart shirts has significantly fewer data points in each segment, which precludes the use of the same architecture. To address this, we experiment with two alternative network configurations for the respiration rate modality: (1) a model consisting of one convolutional layer followed by a GRU layer and a fully connected layer, and (2) a model with two GRU layers followed by two fully connected layers. We find that the model with two GRU layers performed better, and thus, we report the results of this model. 

Additionally, we investigate commonly used statistical features of the respiration rate, including the mean respiration rate during meditation, standard deviation, kurtosis, skewness, entropy, and the longest durations at maximum and minimum respiration rates. Using these features extracted from the Hexoskin data, we employ a Support Vector Machine (SVM) to estimate changes in mindfulness skills.

Figure \ref{fig:mode_all} illustrates that the model utilizing accelerometer data achieves a 15.4\%–21.9\% higher F1 score than the DL model using true respiration rates and a 13.1\%–34.1\% higher F1 score than the SVM model using statistical features derived from respiration rate data. These results demonstrate the superior performance of the accelerometer modality. The accelerometer data not only captures respiration patterns but also encapsulates subtle bodily movements and postural adjustments that are relevant to mindfulness practices. This richer dataset provides additional contextual information indicative of changes in mindfulness skills, leading to more accurate predictions than models using respiration rates alone. The accelerometer inputs offer a more comprehensive representation, enhancing the model’s ability to detect changes in mindfulness skills. 

Additionally, the performance of the SVM model with the statistical feature set demonstrates the quantitative relationship between the respiration biosignal and changes in mindfulness skills. However, the SVM model lacks the capability to capture temporal dynamics in the data, and the limited data points per segment due to lower sampling rates constrain the deep learning model using respiration rate data. In contrast, our proposed DL model using accelerometer data effectively captures both spatial and temporal patterns, resulting in better performance than classical machine learning algorithms like SVM and deep learning models constrained by limited data. 

\parlabel{Assessment of Network Architecture Choices}
We further evaluate the effectiveness of our proposed network architecture by comparing it with EfficientNet, a convolutional neural network (CNN) that has been used in medical image classification and EEG analysis \cite{agarwal2023automated, xiong2021study}. To train EfficientNet on our dataset, we convert the same two-minute accelerometer segments used in our model into mel-spectrograms \cite{davis1980comparison, biswas2023locus}. This conversion allows us to represent the time-series accelerometer data in a format suitable for EfficientNet, which is designed for image-like inputs.

As shown in Figure \ref{fig:net_all}, our proposed model outperforms EfficientNet by achieving a 5.9\%–12.5\% higher F1 score. While the performance of EfficientNet demonstrates that accelerometer data contains sufficient information to identify changes in mindfulness skills, our proposed model achieves better results because it integrates Convolutional Neural Network (CNN) layers with a Gated Recurrent Unit (GRU), effectively capturing both spatial and temporal dependencies in the data.

The superior performance of our model can be attributed to its ability to capture temporal information associated with slow-paced respiration rates, which is crucial for accurately predicting improvements in mindfulness skills. While EfficientNet excels at extracting spatial features from image-like data, it may not effectively model the temporal sequences inherent in accelerometer data related to mindfulness practices. In contrast, our model’s architecture is designed to handle time-series data, leading to better overall performance in this context.

\parlabel{Implication and Conclusion}
Our approach is grounded in both empirical evidence and theoretical frameworks linking physiological signals to mindfulness components. Our deep learning model utilizes accelerometer data captured during meditation sessions to predict changes in mindfulness skills. This choice is informed by research indicating that respiration patterns, detectable via accelerometry, are closely associated with mindfulness states. Specifically, slow and controlled breathing has been shown to activate the parasympathetic nervous system, promoting relaxation and reducing anxiety. These physiological changes are integral to mindfulness practices~\cite{malinowski2013neural}.

Furthermore, our model captures nuanced patterns in the accelerometer data, including breathing rate, variability, and motion artifacts, which reflect potential distractions or shifts in attention. By analyzing these features, the model differentiates between various aspects of mindfulness: concentration, sensory clarity, and equanimity. This aligns with Shinzen Young’s Unified Mindfulness framework \cite{young2016mindfulness}, which defines mindful awareness as the integration of these three components.

Overall, these results highlight the robustness of our approach to estimating changes in mindfulness skills using readily available smartphone data. By accurately predicting improvements in equanimity, concentration, and sensory clarity, our model demonstrates significant potential for facilitating personalized mindfulness training through feedback. This approach paves the way for scalable, accessible mental health interventions that leverage everyday technology. By specifically comparing the accelerometer-based model with models using true respiration rates, statistical features, and alternative network architectures like EfficientNet, we establish that accelerometer data provides a more
comprehensive and informative input for predicting mindfulness skill improvements. The accelerometer captures nuanced bodily movements and temporal patterns that respiration rate data alone cannot fully represent. Moreover, our tailored DL architecture effectively harnesses this rich data, outperforming architectures like EfficientNet that are not optimized for time-series data. Our findings suggest that utilizing accelerometer data in conjunction with advanced DL architectures can significantly enhance the detection of mindfulness skill progression, offering a valuable tool for personalized mental health support.

\section{User study}
In this section, we examine how integrating a respiration biosignal feedback chart into our mindfulness training application influences both system usability and mindfulness skill development. Usability was assessed using the System Usability Scale (SUS), while changes in mindfulness skills were evaluated through session-level feedback.

Participants were randomly assigned to one of two groups: (1) the biosignal-augmented group, which received respiration biofeedback after each session, and (2) the control group, which did not receive any feedback post-session.

The SUS scores and mindfulness skill changes were analyzed using data from Studies 2 and 3, which together included 32 participants. User engagement—measured by the number of completed sessions and total meditation minutes—was evaluated only in Study 3, which involved 22 participants who self-regulated their meditation practice. In contrast, participants in Studies 1 and 2 were required to complete all sessions, limiting natural variability in engagement behavior.

Details of each study's design are provided in Table~\ref{tab:study}, and participant demographics are described in Section~\ref{data collection}.

\subsection{Number of Participants selection from Power Analysis}
We select the number of participants for our user study by conservatively using the effect size of previous studies. There is a wide range of studies that examine the effects of adding gamification or user support features to digital mental health apps on user engagement and performance outcomes \cite{yin2022impact, ciuchita2023really, torous2020dropout} . As might be expected, there’s a wide range of effect sizes for these outcomes (d’s 0.37-1.62). No previous studies to our knowledge have examined the role of respiration feedback to help us more precisely estimate an effect size, but Tsay et al. show that performance is significantly better in their gamified learning courses relative to the nongamified condition (Cohen's d=1.51) and engagement is positively related to performance \cite{tsay2018enhancing}. From this previous literature, we conservatively estimate an effect size of d=1.32 on user engagement in the respiration biosignal feedback augmented condition relative to the standard no feedback condition. A power analysis using G*Power \cite{kang2021sample} for this independent samples t-test, assuming power 0.8 and two-tailed p=.05, indicates that a sample size of at least 22 participants is needed. For our user engagement observation, we use a sample size of 22, and in other studies, we use a sample size of 32, combining the participants from the lab.

\subsection{System Usability Scale}
We assess application usability using the System Usability Scale (SUS) \cite{toker2012job} during the post-study debriefing questionnaires of all studies. The SUS prompted participants to evaluate the usability of the mindfulness meditation app with statements such as: “I think that I would like to use the app frequently,” “I found the app unnecessarily complex,” and “I felt confident using the app.” Responses are given on a scale from 1 (strongly disagree) to 5 (strongly agree). A score of 68 is considered average, while a score of 80 corresponds to an “A” rating \cite{lerner2004unemployment, perciavalle2017role}.

We hypothesize that the biosignal augmented condition (participants who are exposed to the respiration biosignal feedback) would yield higher SUS scores in comparison to the control condition. We consider the participants from studies 2 and 3 for the SUS score. In total, scores from 32 participants (16 participants in the biosignal augmented condition and 16 participants in the control condition) are used to prove our hypothesis.  An independent samples t-test (N=32) reveals that participants in the biosignal augmented condition report significantly higher SUS scores (mean 84.43, SE 1.80) in comparison to those in the control condition (mean 74.04, SE 3.49; t = 2.649; P = 0.013; Cohen’s d = 0.83; 95\% CI =  [2.37,18.43]).

The SUS analysis demonstrates statistically significant results with a p-value <0.05 and a high effect size of 0.83. It implies that incorporating respiration biosignal feedback into the app enhances user satisfaction and perceived usability. Participants reported higher satisfaction and confidence when using the app with biosignal feedback, suggesting that this feature contributes positively to the overall user experience. Improved usability is crucial for sustained engagement with mindfulness applications, ultimately supporting better mindfulness skill development, which has been shown to improve mental health outcomes.

\parlabel{In-Lab Study Results}
Ten participants from study 2 who completed their sessions in the lab are divided into experimental and control conditions. An independent samples t-test using all available data collected in lab environment (study 2) revealed significantly higher user satisfaction in the biosignal augmented condition on the SUS (Mean=85.0, SE=2.11) relative to control condition on the SUS (Mean=78.57, SE=1.99), t = -2.21, p =.04, Cohen's d = 0.67, 95\% confidence interval (CI) = [-1.91, - 0.04]. While the sample size is small, the effect size (Cohen's d) indicates that participants may be more satisfied with their training experience when exposed to respiration biosignal feedback.

\parlabel{In-The-Wild Study Results}
Similar to the lab study, an independent samples t-test using all available data from study 3 revealed significantly higher user satisfaction in the biosignal augmented condition on the SUS (Mean=84.36, SE=3.83) relative to control condition on the SUS (Mean=67.08, SE=8.52), t = 2.02, p = 0.020, Cohen's d = 1.04, 95\% confidence interval (CI) = [-1.31, 6.96]. The result is statistically significant with a p-value < 0.05 and has a high effect size of 1.04 (Cohen's d), which indicates that exposure to respiration biosignal feedback increases participants' satisfaction in the wild.

\subsection{Impact on Mindfulness Skill Change}
\begin{figure*}[t!]
    \centering
    \begin{subfigure}[t]{0.35\textwidth}
        \centering \includegraphics[width=\textwidth]{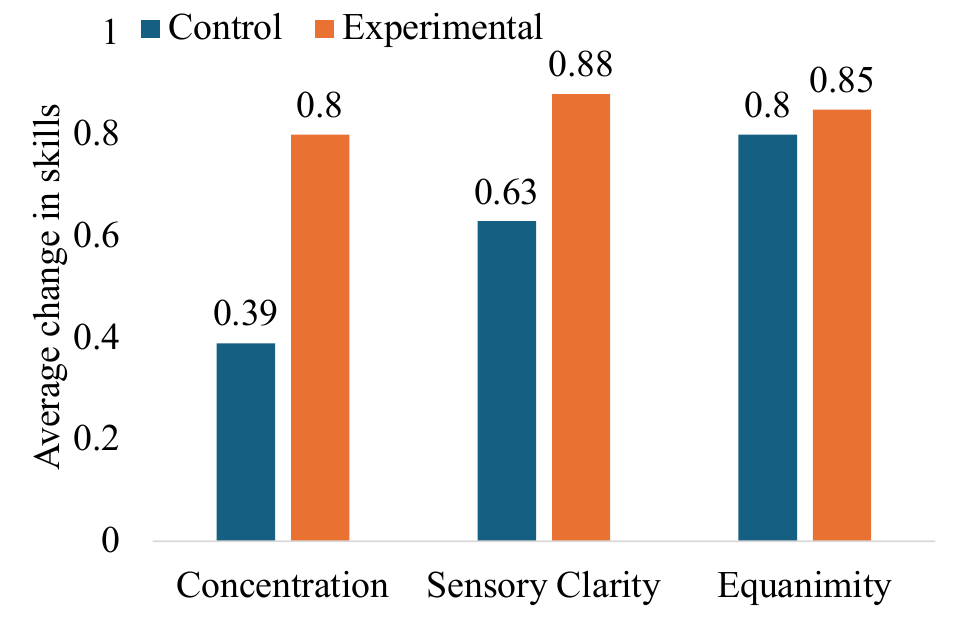}
        \caption{Average change in mindfulness skills}
        \label{fig:study3_change}
    \end{subfigure}%
    \hspace{0.5em}
    \begin{subfigure}[t]{0.62\textwidth}
        \centering
        \includegraphics[height=1.4in]{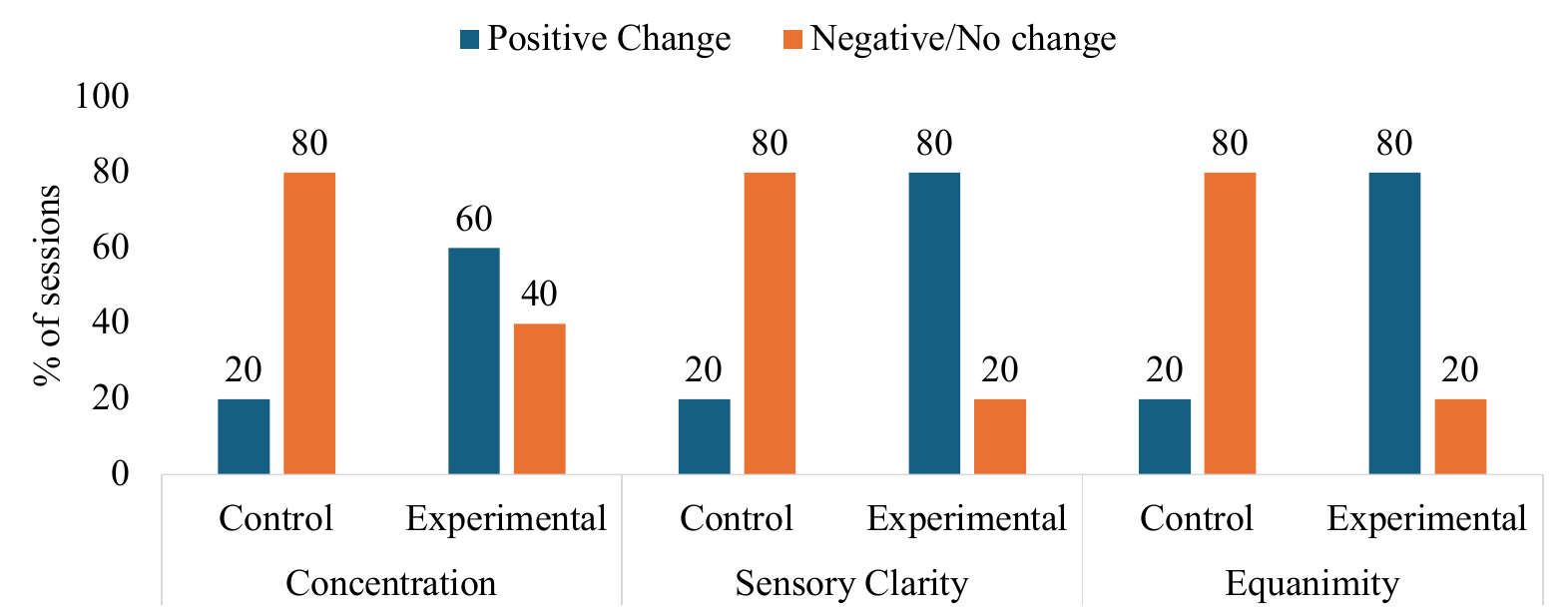}
        \caption{Percentage of session with positive changes}
        \label{fig:study3_progress}
    \end{subfigure}
    \caption{Impact of respiration biosignal feedback shows higher skill change and higher number of sessions with positive changes in the experimental group than the control group.}
\end{figure*}
We evaluate the impact of the respiration biosignal feedback chart on changes in mindfulness skills following meditation. Figure \ref{fig:study3_change} shows the average change in mindfulness skills for both the control and experimental conditions. Participants who receive respiration biosignal feedback exhibit up to 2 times greater improvement in mindfulness skills compared to the control group. In addition, Figure \ref{fig:study3_progress} indicates that the experimental group experiences more sessions with a positive change in mindfulness skills. These findings suggest that respiration biosignal feedback not only enhances app usability but also promotes ongoing improvements in mindfulness skills as sessions progress.

Additionally, we conduct a t-test on the average change in mindfulness skills and a chi-square test on the number of positive changes in mindfulness skills due to its categorical nature. The average change in concentration and sensory clarity is highly significant with a large effect size, resulting in a p-value of 0.0006 and 0.026, respectively. However, the average change in equanimity is not statistically significant, with a p-value of  0.64 and Cohen’s d of 0.17. The change in concentration and sensory clarity also has a large effect size with Cohen's d of 1.37 and 0.83, respectively, while the change in equanimity has a low effect size of 0.17.

On the contrary, the chi-square test shows that the number of sessions with positive changes in skills is highly significant for all three skills of concentration, sensory clarity, and equanimity, with a p-value of $1.8 \times 10^{-8}$, $7.2 \times 10^{-7}$, and $7.2 \times 10^{-7}$ and a Cohen’s h of -0.86, -1.29, -1.29, respectively, showing large effect size.

\subsection{User Engagement}
\begin{figure}
    \centering
    \includegraphics[width=0.8\linewidth]{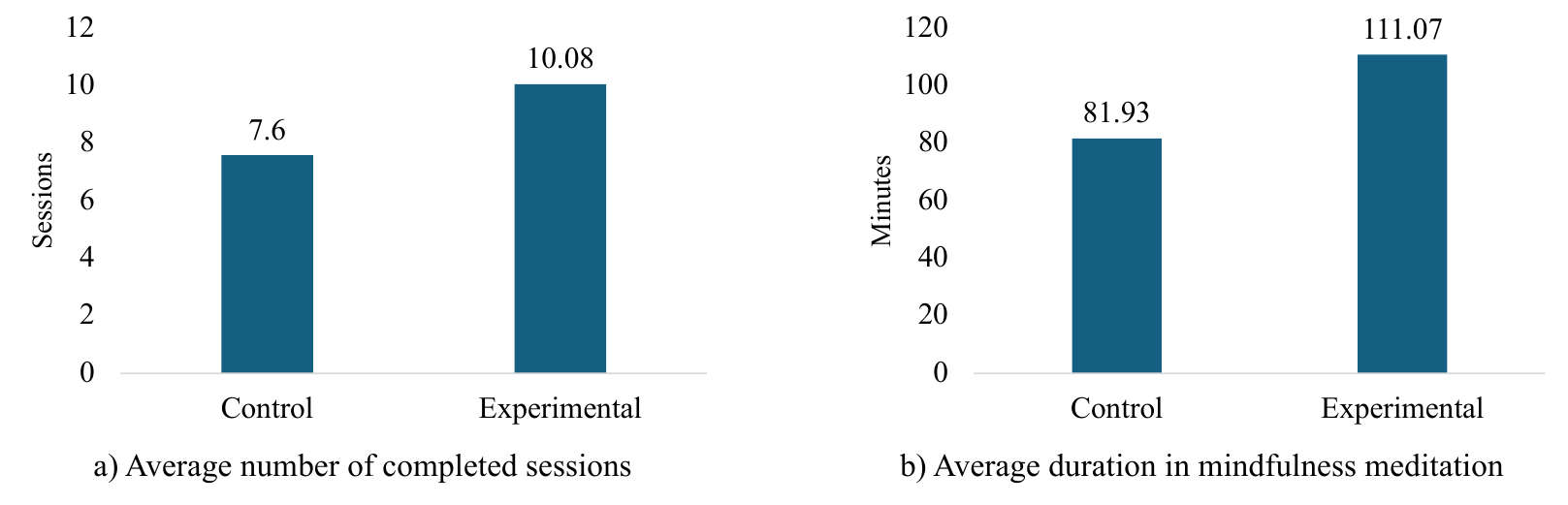}
    \caption{Participants in the biosignal-augmented condition exhibited higher engagement with the app, completing more sessions on average and spending more time in meditation.}
    \label{fig:user_engagement}
\end{figure}
In addition to assessing system usability, we also explore user engagement by providing an experimental group of users with respiration biosignal feedback. Because all participants in Studies 1 and 2 were required to visit the lab, we do not track their engagement data, as it would not accurately reflect real-world usage patterns. As mentioned in section \ref{data collection}, participants from study 3 are divided into a biosignal augmented condition and a control group, and they can complete as many sessions as they want in 21 days.

As shown in Figure \ref{fig:user_engagement}, participants in the in-the-wild study who received biosignal-augmented feedback completed an average of 10.08 meditation sessions, while the control group averaged 7.60 sessions. 
Those in the biosignal-augmented condition spent more time meditating, averaging 111.07 minutes, while the control group averaged 81.93 minutes. These trends highlight the potential of respiration biosignal feedback to enhance engagement and encourage longer meditation sessions. 
A t-test (N=22) proves the results are approaching significance, with a p-value of 0.11. The effect size is medium to large with a Cohen’s d of 0.71. A larger longitudinal study with more participants will reinforce our hypothesis.

%% file: IMWUT_25/tex/06_discussion.tex
\section{Discussion, Limitation, and Future Works}
This section highlights several key limitations and areas for improvement in our study, including using a custom mindfulness scale and binary classification for skill detection. We also address challenges involving sample size, device dependence, and environmental artifacts, all of which underscore the need for further investigation and methodological refinements.

\subsection{Scale for Mindfulness Skills} 
One limitation of our study is the use of a custom 1–7 scale for participants to self-report their current levels of concentration, sensory clarity, and equanimity. While this scale provided a concise method for capturing mindfulness skills, it differs from standard validated scales commonly used in mindfulness research. Due to the necessary brevity of our self-report questionnaires and considering that previous studies \cite{lindsay2019mindfulness, lindsay2018mindfulness} indicate increases in self-reported mindfulness measures even in control conditions not exposed to mindfulness training \cite{park2013mindfulness}, we opt for this concise scale. In future work, we plan to incorporate more widely used and validated scales, such as the Mindful Attention Awareness Scale (MAAS) \cite{mackillop2007further} and the Toronto Mindfulness Scale (TMS) \cite{lau2006toronto}, to enhance the validity and comparability of our findings. 

\subsection{Ground Truth and Practicality of Mindfulness Assessment}
In this study, we use self-reported scores as the ground truth for training our mindfulness skills estimation model. While self-report measures are subject to potential biases such as mood variability or subjective interpretation, they remain the most practical and widely accepted method for evaluating mindfulness experiences in large-scale, real-world studies.

More objective physiological assessments—such as EEG or fMRI—can offer higher precision in controlled laboratory settings. However, these modalities are often cost-prohibitive, lack scalability, and may disrupt the natural meditation experience. As such, they are not well-suited for mobile, in-the-wild mindfulness interventions. Our approach aims to balance ecological validity and practical deployment by leveraging self-reports alongside non-intrusive, smartphone-based sensing methods.

We acknowledge that self-reported scores alone may not fully capture the complexity of mindfulness. To address this, future work will explore the integration of additional physiological markers, such as heart rate variability (HRV) and galvanic skin response (GSR), to enable multi-modal validation. These extensions will help enhance the robustness and generalizability of our mindfulness assessment framework in naturalistic settings.

\subsection{Granularity of Mindfulness Skill Prediction} 
Our proposed model currently focuses on binary classification— predicting whether mindfulness skills are improving or not during training. This binary approach, while useful, may not capture the full spectrum of changes in mindfulness skills. In future work, we aim to extend the model to a multi-class classification framework that can quantify varying degrees of change, such as large positive change, low positive change, no change, and negative change in concentration, sensory clarity, and equanimity. By providing more granular predictions, we can offer users a deeper understanding of their mindfulness progression, potentially enhancing the personalization and effectiveness of the training. 

\subsection{Limited Scope of Respiratory Biosignals} 
While our proposed model for estimating respiration rate achieves an MAE of 1.6 BPM, outperforming state-of-the-art models, it focuses solely on respiration rate estimation. Future research will aim to develop a more robust model to further improve the accuracy of respiration rate estimation, particularly under varying conditions. Moreover, we plan to expand our focus to estimate additional respiratory biomarkers, such as minute ventilation, inhalation and exhalation durations, and respiratory variability. These biomarkers are vital for comprehensive monitoring of user health and well-being \cite{bari2020automated, plarre2011continuous}, and their inclusion could enhance the biofeedback provided to users during mindfulness training. 

\subsection{Exploring Alternative Modalities for Respiration and Mindfulness Estimation}
While this study leverages smartphone accelerometer data to estimate respiration rate and track changes in mindfulness skills during meditation, other physiological sensing modalities—such as EEG, audio, and galvanic skin response (GSR)—were not explored. Recent work has demonstrated the feasibility of estimating respiration from in-ear audio signals~\cite{breathtrack, breeze, ahmed2023remote}. However, capturing respiration using external microphones presents additional challenges, particularly in mindfulness contexts where breathing is slow, soft, and often inaudible. This makes external audio sensing less reliable for detecting low breathing rates in real-world scenarios.

While in-ear microphones may have the capability to capture subtle breathing signals, microphone data access from commercial earbuds (e.g., Samsung Galaxy Buds, Apple AirPods) is restricted. Moreover, although these devices also include embedded IMUs, their sensor data are not publicly accessible—limiting their applicability for research and large-scale deployment. Similarly, EEG and GSR, though capable of providing rich physiological insights, require external sensors and more complex setups, reducing their practicality for everyday use.

In future work, we aim to explore the potential of external microphone audio to estimate both respiration rate and mindfulness skills, while carefully evaluating its limitations under low-breathing-rate conditions. These efforts will help determine whether complementary sensing modalities can enhance estimation accuracy without compromising usability or scalability.

\subsection{Long-Term Engagement and Future Directions}
Our findings indicate that integrating the respiration feedback module enhances user engagement, as evidenced by an increase in both the number of meditation sessions and the total minutes spent meditating. These results suggest that providing real-time respiration biofeedback and mindfulness skill estimation may positively influence user commitment to mindfulness practices. While this trend is promising and statistically significant, the current study was conducted over a 21-day period and represents an initial step toward understanding engagement dynamics.

To evaluate the sustainability of this effect, a longitudinal study is necessary. In future work, we plan to conduct an extended 6–12 month deployment with a larger and more diverse participant pool. This will allow us to assess long-term user engagement and determine whether the initial increase in usage is maintained over time. Additionally, future studies will incorporate more controlled experimental designs to better isolate the impact of biosignal feedback from other confounding variables. These efforts will help strengthen the evidence for the role of physiological feedback in enhancing digital mindfulness interventions and inform the design of scalable, personalized mindfulness tools.

\subsection{Sample Size and Participant Diversity} 
A limitation of our study is the relatively small sample size, which includes only 261 sessions. While our findings are promising, the limited number of participants may affect the generalizability of the results. Future studies should include a larger sample to validate the findings across different populations, ages, and backgrounds.

\subsection{Device and Sensor Limitations} 
Our respiration rate estimation algorithm and mindfulness skill detection model are developed using data collected from specific devices, namely the iPhone X and Hexoskin smart shirts. The reliance on these specific devices may limit the generalizability of our methods to other smartphones or wearable devices with different sensor characteristics. Future work should explore the applicability of our algorithms across a wider range of devices and platforms to enhance their usability and accessibility.

\subsection{Environmental and Movement Artifacts} 
Although our algorithm includes steps for motion artifact removal, continuous user movement or improper device placement can still impact the accuracy of respiration rate estimation and mindfulness skill detection. In real-world settings, uncontrolled environmental factors and varying levels of user compliance may introduce additional noise and artifacts. Future research should focus on enhancing the robustness of the algorithms to handle such variability, possibly through adaptive filtering techniques or machine learning models that can account for diverse real-world conditions.


%% file: IMWUT_25/tex/07_conclusion.tex
\section{Conclusion}
This study demonstrates the significant potential of integrating respiration biosignal feedback into smartphone-based mindfulness training apps. Our respiration rate estimation algorithm outperformed existing methods in both accuracy and reliability, particularly excelling at detecting slow respiration rates typical during mindfulness meditation. By maintaining a low MAE of 1.6 for various respiration rates, our algorithm ensures effective biofeedback, which is crucial for mindfulness practices. Additionally, our deep learning model for mindfulness skill estimation using accelerometer data achieved robust performance, with F1 scores of 82\% for concentration, 80\% for sensory clarity, and 84\% for equanimity. Our model surpassed other models that are based solely on respiration rate, highlighting the value of accelerometer data in capturing subtle physiological and behavioral cues associated with mindfulness skills. Participants who receive biosignal feedback reported higher satisfaction with using the app. These findings suggest that incorporating biosignal feedback, combined with accelerometer-based mindfulness estimation, can significantly enhance the effectiveness of digital mindfulness interventions. By harnessing ubiquitous technology and sophisticated data analysis, we can create effective, scalable solutions to support mindfulness practice and mental health in diverse populations.

\section{Acknowledgments}
Research reported in this publication is supported by the National Institute of Mental Health of the National Institutes of Health under Award Number R44MH134709. The content is solely the responsibility of the authors and does not necessarily represent the official views of the National Institutes of Health. We would like to acknowledge the contributions of Julianna Raye and Chad Frisk, who designed and developed the curriculum used in the Equa app.